\documentclass[12pt]{article}
\usepackage{graphicx}

\begin{document}

\title{ {\bf Four-point functions in Keldysh basis}\thanks{Preprint INLN 2001/12 }}
\author{F. Guerin \\ {\it Institut Non Lineaire de Nice}  \\  {\it 1361 route des Lucioles,  06560 Valbonne, France}}
\date{}
\maketitle

\begin{abstract}

The Keldysh three-and four-point functions are related to analytical continuations of the Imaginary Time amplitudes, in momentum space. The relation provides an interesting interpretation of the amplitudes that are specific to the Keldysh basis, and of the Ward identities they satisfy in QED. The interpretation carries over to the (perturbative) non-equilibrium case. 

\end{abstract}


\section{Introduction}
For out of equilibrium situations, the only known formalism for perturbative calculations of Green functions is the Complex-Time-Path (CTP) formalism (or Keldysh formalism) \cite{Kel,Chou}. This is in contrast to the thermal equilibrium case where there exists several formalisms which are completely equivalent (Imaginary-Time (IT) \cite{MLB}, Real-Time $1/2$ \cite{MLB}, Retarded-Advanced ($R/A$) \cite{Au1}, Keldysh  \cite{Chou})

In a series of papers \cite{Ca3,Ca2,Ca1,Ca4,Ca5}, Carrington and collaborators have gained experience in the Keldysh formalism for 2-point, 3-point, 4-point functions  in thermal equilibrium, together with some generalization to non-equilibrium situations. The Keldysh formalism is very opaque, and it  is so cumbersome that a Mathematica program \cite{Ca6} has to be used to write down one-loop diagrams'  integrands.

In the paper "Four-point spectral functions and Ward identities in hot QED" \cite{Ca1} Hou Defou, Carrington, Kobes, and Heinz consider three 4-point functions that are specific to the Keldysh basis, in addition to the well known 4-point retarded amplitudes. With the help of the Mathematica program, they compute their one-loop expressions in QED for the "hard-loop" case, in and out of equilibrium. For those Keldysh amplitudes, they obtain forms where the thermal electron mass (or its generalization) is multiplied by overall factors $c_{\beta},d_{\beta}$ which are respectively zero and $\ln 2$ in the thermal equilibrium case.  This result raises questions. Indeed, in thermal equilibrium, any real-time $n$-point amplitude is either an analytic continuation of the unique Imaginary-Time amplitude, or a linear combination of these continuations \cite{Ev}. Therefore: i) how can an overall $\ln 2$ factor  be generated from this linear combination in the "hard-loop" limit ? ii) how is it compatible with the Ward identities for this 2-photon-2-fermion amplitude, which relate 4-point to 3-point, as well as 3-point to the 2-point  Keldysh  fermion self-energy ?

In thermal equilibrium, the equivalence between the three real-time formalisms arises in perturbation theory from  the existence of Bogoliubov-like transformations between the three bases $1/2$, $R/A$, Keldysh \cite{Au1, vW, vWK}.  In this paper, we use the one-loop QED case in order to establish the relations between the 3-point and 4-point QED Keldysh amplitudes and the $R/A$ amplitudes. Those are general relations and they have to be satisfied in this simple example. Moreover, the $R/A$ amplitudes are well specified analytical continuations of the IT amplitudes. A nice outcome is that the relation provides an interesting interpretation of the Keldysh amplitudes in terms of discontinuity in an external energy  (i.e. in terms of cut diagrams). This interpretation is generally valid in thermal equilibrium.  It generalizes to the one-hard-loop out-of-equilibrium case, and very likely to all orders.  \\
This paper stays in the framework defined by the authors of Ref.\cite{Ca1}  for their proof of the Ward identities in the non-equilibrium case :
\\ i) the out-of-equilibrium case is perturbative, i.e. bare massless particles have  out-of-local-equilibrium distribution functions $n(k)$ instead of the thermal ones $n_B(k), n_F(k)$, 
\\ ii) the "hard-loop" QED approximation is restricted to the factor that involves the $\gamma$ matrices, i.e. in this factor: a) neglect all external momenta compared to the loop momentum $K$, b) ignore factors of $K^2$. \\(For the hard-thermal-loop effective theory, this step is only the first one in a series of approximations \cite{MLB}). This approximation is welcome. All the spin part is  carried by the loop momentum, so that all the relations to be established are simple algebraic identities between scalar integrands.

This paper is self-contained in the following sense. The initial starting point for one-loop integrands of the 3-point and 4-point amplitudes were obtained by the authors of Ref.\cite{Ca1} by means of the Mathematica program. In fact, the resulting integrands may be written down by means of simple diagramatic rules (to be justified later on). A reader with paper and pencil may check all the equations. 
Before addressing the 4-point QED functions, time will be spent on the 3-point QED functions. Indeed many features show up in this simpler case: i) the relation to analytical continuations of the IT amplitudes, ii) the interesting interpretation of the Keldysh amplitudes, iii) the interpretation of their Ward identities. Most of the discussions will take place in the equilibrium case. However, this case puts such strong constraints on the Keldysh basis that there seems to be little leeway in the generalization to the (perturbative) non-equilibrium case. \\
The rules for the diagrams' integrand are in Sec.\ref{se2}. In Sec.\ref{se3} the 3-point amplitudes are studied. Sec.\ref{se4} is devoted to the 4-point case, i.e. one loop results, their interpretation, a review of the IT analytical continuations and their relations to the Keldysh amplitudes. Conclusions are in Sec.\ref{se5}. An appendix discusses the interpretation of the 4-point Ward identities.

\section{Diagrams' rules}
\label{se2}
\subsection{The notation}
\label{se2.1}
The notation is that of Ref.\cite{Ca1}. \\
To a propagator carrying momentum $K_i$ are associated
\begin{equation}
r_i=D_R(K_i) \ \ \ ,\ \ \ a_i=D_A(K_i) \ \ \ , \ \ \ f_i=D_F(K_i)
\label{1.1}
\end{equation}
where $D_R, \  D_A$ are the retarded, advanced propagators and $D_F$ is the propagator's component specific to the Keldysh basis.
\begin{equation}
D_R(K)= {1\over{(k_0+i\epsilon)^2-{\mathbf k}^2}} 
 \ \ , \ \  D_A(K)={1\over{(k_0-i\epsilon)^2-{\mathbf k}^2}}
\label{1.2}
\end{equation}
\begin{equation}
D_R(K)-D_A(K) = \ - 2\pi i \ {\rm sgn}(k_0) \ \delta(K^2)
\label{1.3}
\end{equation}
\begin{equation}
D_F(K) = N(K) \ (D_R(K)- D_A(K))
\label{1.4}
\end{equation}
In thermal equilibrium, respectively for bosons and fermions, $N(K)$ is
\begin{eqnarray}
N_B(K)=N_B(k_0)=(1+2n_B(|k_0|)) {\rm sgn}(k_0) =\coth {\beta\over 2}k_0  \nonumber \\
N_F(K)=N_F(k_0) = (1-2n_F(|k_0|)) {\rm sgn}(k_0) =\tanh {\beta\over 2}k_0
\label{1.5}
\end{eqnarray}
with $\beta=1/T$. A property to be used frequently is
\begin{equation}
N(K)=-\ N(-K)
\label{1.6}
\end{equation}
The figures of Ref.\cite{Ca1} associated with the diagrams to be considered are reproduced on Figs.\ref{fi1},\ref{fi2}. $P_1$ is the incoming fermion momentum, $- P_2$ the outgoing fermion momentum, $P_3, \  P_4$ are incoming photon momenta. In the loop, $K_1=K$ is always the photon momentum, $K_2, \ K_3, \ K_4$ are fermion momenta, i.e. $1$ always refer to a boson propagator, $2,3,4$ to fermion propagators.

\subsection{The rules}
\label{se2.2}
The scalar part of the one-loop integrands for the 3-point and 4-point one-particle-irreducible Keldysh functions have been obtained in Ref.\cite{Ca1,Ca2} by the summation of (many) terms, each one obtained from the contraction of 1/2 indices (the indices correspond to the two indices of the CTP contour). This is a very cumbersome task already for the 3-point one-loop integrand, as a propagator is a two-index tensor and a vertex is a three-index tensor, and so a Mathematica program was used.  It turns out that the resulting integrands can be written down if one follows simple diagramatic rules. For the $n$-point retarded amplitudes, those rules are known and they are easily justified (set ${\cal A}$ of rules). For the amplitudes specific to the Keldysh basis that are considered in Ref.\cite{Ca1}, other rules will be stated (set ${\cal B}$ of rules) and their justification will appear later on.
The rules involve the quantities $r_i, \ a_i, \ f_i, \ N(K_i) $ associated with a propagator of momentum $K_i$ and defined in Eqs.(\ref{1.1})(\ref{1.4}).
\\
{\it \bf Set ${\cal A}$ of rules} (for retarded amplitudes) \\
(i) All external momenta are incoming. Each external momentum carries an index $R$ or $A$. A momentum $P_j$ is $(p_j^0+i\epsilon_j, {\mathbf p})$. Then $p_{j R}$ corresponds to $\epsilon_j>0$,  $p_{jA}$ to $\epsilon_j<0$. In an $n$-point amplitude $\sum_{j=1}^n P_j=0$ and  $\sum_{j=1}^n \epsilon_j=0$. \\
(ii) A one-loop diagram  may be written as the sum of terms obtained by putting successively on-shell one of the loop propagators, i.e. it is the sum over all possible ways of cutting once the loop so that a connected tree diagram is obtained. \\
(iii) To the cut propagator $K_i$, one associates a factor
$ f_i\ = \ N(K_i)  \  (r_i-a_i) \ = \ N_i \ (r_i-a_i) $. \\ 
(iv) One can define an $\epsilon$-flow along each tree. The flow is entering the external lines of type $R$, and outgoing the external lines of type $A$. Along the tree, the $\epsilon$-flow obeys the rules of an electric current at each vertex. The flow ignores the cut line(s). \\
(v) The loop momenta are oriented anticlockwise. To a tree propagator carrying the momentum $K_j$ one associates $r_j$ if the loop momentum is parallel to the $\epsilon$-flow, and $a_j$ if it is antiparallel. 

For the thermal equilibrium case, those rules just correspond to picking up the poles of the integrand when deforming the $k_0$ integration contour from the imaginary axis to the real axis (the difference between the $T=0$ and $T\neq 0$ cases lies in the weight associated with the pole's residue). The $\epsilon$-flow keeps track on the $\epsilon$ prescription that exists on the external lines \cite{FG1}.
\\
{\it \bf Set ${\cal B}$ of rules } (for a set of Keldysh amplitudes) \\
(i) The external momenta are defined as in rule (i) of set ${\cal A}$. \\
(ii) The momentum running around the loop is cut twice, i.e. the diagram is cut into two tree-like structures. \\
(iii) To the cut lines $i$ and $j$ one associates
$ f_i  f_j  -  (r_i-a_i)(r_j-a_j) \ = \ (N_iN_j - 1) (r_i-a_i)(r_j-a_j) $. \\
(iv) The $\epsilon$-flow along each tree obeys the rule (iv) of set ${\cal A}$. \\
(v) To a tree's propagator of momentum $K_j$ one associates  a factor $r_j$ or $a_j$ according to the rule (v) of set ${\cal A}$. \\
(vi) The allowed cuts into two trees are those where the outgoing $\epsilon$-flow is connected to an incoming $\epsilon$-flow. In contrast, an incoming $\epsilon$-flow may have no exit. \\
(vii) One sums over all allowed cuts. \\
(viii) The rule (vi) does not apply to the 2-point case.

\section{Two- and Three-point amplitudes}
\label{se3}
\subsection{The two-point amplitudes}
\label{se3.1}
This case serves as a reminder of useful facts. \\
The one-loop fermion self-energy is drawn on Fig.\ref{fi1}a. $P$ is the incoming fermion momentum, $K=K_1$ refers to the photon momentum in the loop, $K_2=K_1-P$  to the fermion loop momentum. With the notations of Sec.\ref{se2.1}, and the approximation $P\ll K$ in the factor involving the $\gamma$ matrices, the two-point amplitudes may be written \cite{Ca2}
\begin{equation}
\Sigma(P) = i e^2\int \ {d_4K\over{(2\pi)^4}} \ \ {\gamma}.K \ \ \tilde{\Sigma} (P; K)
\label{2.1}
\end{equation}
The set ${\cal A}$ of rules give the one-loop retarded amplitude
\begin{equation}
\tilde{\Sigma}_R(P;K) = f_1a_2+f_2r_1
\label{2.2a}
\end{equation}
$\Sigma_A(P)=\Sigma_R^*(P)$ so that the corresponding integrand is
\begin{equation}
\tilde{\Sigma}_A(P;K) = f_1r_2+f_2a_1
\label{2.2b}
\end{equation}
\begin{equation}
\tilde{\Sigma}_R-\tilde{\Sigma}_A = f_2(r_1-a_1)-f_1(r_2-a_2) = (N_2-N_1)(r_1-a_1)(r_2-a_2)
\label{2.3}
\end{equation}
The set ${\cal B}$ of rules give the one-loop Keldysh amplitude
\begin{equation}
\tilde{\Sigma}_F(P;K) = f_1f_2 -(r_1-a_1)(r_2-a_2) =(N_2N_1-1)(r_1-a_1)(r_2-a_2)
\label{2.4}
\end{equation}
\begin{equation}
\tilde{\Sigma}_F = f_1f_2+r_1a_2+r_2a_1
\label{2.5}
\end{equation}
Two comments follow:

In the transition from Eq.(\ref{2.4}) to Eq.(\ref{2.5}), a term $r_1r_2+a_1a_2$ has been dropped, with the argument that it gives a vanishing contribution. Indeed, in the integration over the energy $k_0$, all the integrand's poles of $r_1r_2$ (or $a_1a_2$) sit on the same side of the real $k_0$ axis, and one may close the contour on the other side. This argument will be a recurrent one in the comparison between our rules' results and the results of Ref.\cite{Ca1}. One should be aware not to introduce ultraviolet divergences in the process. $\Sigma_F$ is a convergent integral when form (\ref{2.4}) is used for $\tilde{\Sigma}_F$.

$\Sigma_R-\Sigma_A$ is the imaginary part of the fermion self-energy. The particles in the intermediate state $(1, -2)$ are on-shell. Both $\Sigma_R-\Sigma_A$ and $\Sigma_F$ are associated with the same intermediate state, however the integrand carries a different weight in both cases, $(N_1-N_2)$ for $\Sigma_R-\Sigma_A$ and $(1-N_1N_2)$ for $\Sigma_F$. 
\\

In the case of {\it thermal equilibrium}, one may go a step further. One makes use of one characteristic properties of the thermal weights (i.e. of the $\coth$ and $\tanh$ functions)
\begin{eqnarray}
1-N_F(K_2)N_B(K_1) \ = \  N_F(K_1-K_2) \ [N_B(K_1)-N_F(K_2)] \nonumber \\
1-N_F(K_2)N_F(K_3) \ = \  N_B(K_3-K_2) \ [N_F(K_3)-N_F(K_2)]
\label{2.7}
\end{eqnarray}
or, in our notations
\begin{equation}
f_if_j-(r_i-a_i)(r_j-a_j) \ = \ N(K_j-K_i) \  [f_i(r_j-a_j)-f_j(r_i-a_i)]
\label{2.8}
\end{equation}
Equation (\ref{2.8}) allows to relate Eqs.(\ref{2.3}) and (\ref{2.4})
\begin{equation}
\tilde{\Sigma}_F(P;K)=N_F(K_1-K_2) \ (\tilde{\Sigma}_R-\tilde{\Sigma}_A)
\label{2.9}
\end{equation}
\begin{equation}
\Sigma_F(P) \ = \ N_F(P) \ ( \ \Sigma_R(P)-\Sigma_A(P))
\label{2.10}
\end{equation}
Relation (\ref{2.10}) is a general relation and the one-loop example has to verify it. A relation similar to Eq.(\ref{2.10}) holds for the scalar field case, where $N_F(P)$ is replaced by $N_B(P)$. Relation (\ref{2.10}) may be interpreted in several ways. One interpretation that will turn out to be useful follows.  It may be seen as the first consequence of the existence of Bogoliubov-like transformations that relate the real time formalisms, i.e. the $1/2$ basis, the $R/A$ basis, the Keldysh basis \cite{Au1, vW, vWK}. The matrix transformation $U$ relating the Keldysh and the $R/A$ matrix propagators, $D'=U(P)  D(P) U^T(-P)$ involves $N_B(P)$ for a boson field, $N_F(P)$ for a fermion field, and it commutes with $\gamma$ matrices \cite{vWK}. Since in the $R/A$ basis, there are only two non zero matrix elements $D_R$ and $D_A$, i.e. two self energy $\Sigma_R$,  $\Sigma_A$, there must be a relation between the Keldysh component $\Sigma_F$ and $\Sigma_R , \Sigma_A$.

Another interpretation of relation (\ref{2.10}) follows. For the pure boson case, the statistical weight of an $n$-particle intermediate state corresponds respectively  to  "emission $-$ absorption''  in $\Sigma_R-\Sigma_A$ and to "emission $+$ absorption'' in $\Sigma_F$  (for an incoming fermion the signs are reversed). Indeed there exists a relation valid for any $n$ . With all momenta outgoing in the intermediate  $n$-particle state, $P=\sum_{i=1}^n K_i $ and $N_i=1+2 n_B(K_i)$ or $N_i=1-2 n_F(K_i)$, the relation is, for example for $n=3$,
\begin{eqnarray}
N(1+2+3)[(N_1+1)(N_2+1)(N_3+1) \ - \ (N_1-1)(N_2-1)(N_3-1)]  \nonumber  \\
= (N_1+1)(N_2+1)(N_3+1) \  + \  (N_1-1)(N_2-1)(N_3-1)
\end{eqnarray}
i.e. the multiplication by $N(1+2)$ transforms $N_1+N_2$ into $1+N_1N_2$, and the multiplication by $N(1+2+3)$ transforms the statistical weight $1+N_1N_2+N_2N_3+N_1N_3$ into $N_1N_2N_3 +N_1+N_2+N_3$. This feature is related to the gemeral properties of the full propagator's components. For bosons, $D_R-D_A$ is the expectation value of the boson field commutator, while $D_F$ is the expectation value of the field anticommutator (the signs are reversed for a fermion field). Hence, one can predict what the Keldysh formalism will give in $\tilde{\Sigma}_F$ for a three-particle intermediate state, in or out-of equilibrium, (such as a two-loop self-energy diagram in $\Phi^4$), it is
\begin{eqnarray}
(N_1N_2N_3 + N_1+N_2+N_3)(r_1-a_1)(r_2-a_2)(r_3-a_3) \nonumber \\
= f_1f_2f_3 \ + \ f_1(r_2-a_2)(r_3-a_3) +f_2(r_1-a_1)(r_3-a_3)+f_3(r_2-a_2)(r_1-a_1)    
\end{eqnarray}
One last comment. $\Sigma_R(P)-\Sigma_A(P)$ is twice the imaginary part of the self-energy $\Sigma(P)$. In what follows, it will be more general to speak of the diagram's discontinuity in the energy variable $p^0$. At finite temperature, a diagram such as the one in Fig.\ref{fi1}a , has branch cuts all along the real $p^0$ axis, and the discontinuity in the complex $p^0$ plane is the difference between the two ways of approaching the real axis, from above $p^0+i\epsilon$, and from below $p^0-i\epsilon$
\begin{eqnarray}
\Sigma_R(P)-\Sigma_A(P) &=& \Sigma(p^0+i\epsilon, {\bf p}) - \Sigma(p^0-i\epsilon, {\bf p})  \ \ , \ \ \epsilon>0
\nonumber \\
\Sigma(p_R) \ - \ \Sigma(p_A) \ &=& \ {\mathrm Disc}_p \ \Sigma(p)
\label{2.6}
\end{eqnarray}
For a given diagram, the contribution to $\Sigma_R-\Sigma_A$ obeys the Cutkosky rules: i) all the particles in the intermediate state are on-shell (i.e. the diagram is cut into two pieces),  ii) the intermediate states are the familiar ones at $T=0$, iii) the weight of the intermediate state has been discussed here up, iv) one sums over all possible intermediate states present in the diagram.

\subsection{The three-point amplitudes}
\label{se3.2}
\subsubsection{One-loop integrands}
\label{se3.2.1}
\begin{figure}[p]
\includegraphics*[width=10cm]{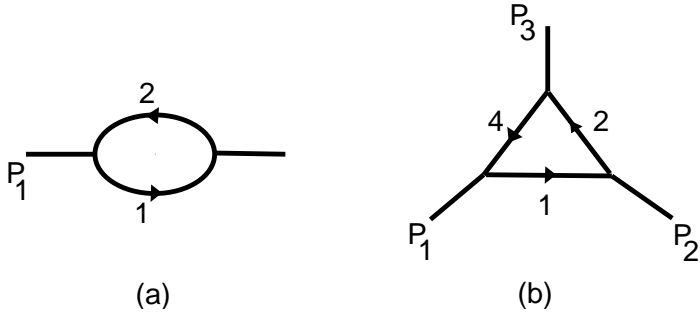}
\caption{\label{fi1} One-loop diagrams: (a) fermion self-energy,  (b) vertex in QED. $P_1$ is the incoming fermion momentum, $P_3$ the incoming photon momentum.  1, 2, 4 refer to momenta in the loop, 1 is a photon,  the other ones are fermions.}
\end{figure}

The one-loop diagram is drawn on Fig.\ref{fi1}b . The  fermion momentum is $P_1$, and $P_3$ is the photon momenta. All momenta are incoming, i.e. $P_1+P_2+P_3=0$. The loop momenta  are $K_1, K_2, K_4$, where $K_1=K$ is the photon momentum and they are oriented anticlockwise.
\begin{equation}
G^{\mu}(P_1, P_2, P_3)=e^3 \ \int \ {d_4K\over(2\pi)^4} \ 2 K^{\mu} \ {\gamma}.K \ \ \tilde{G}(P_1, P_2, P_3; K)
\label{2.11}
\end{equation}
(the addition to the bare vertex  reads $ie\gamma^{\mu} + G^{\mu}$). The retarded amplitudes are obtained from the set ${\cal A}$ of rules. Only one  external leg is of type $R$. There are three ways to cut once the loop momentum, i.e. the integrand is the sum of three terms
\begin{eqnarray}
\tilde{G}_R &=& \tilde{G}_R(p_{1A},p_{2R},p_{3A}\ ; K) = f_1r_2r_4+f_2a_1a_4+f_4a_1r_2  \nonumber \\
\tilde{G}_{Ri} &=& \tilde{G}_R(p_{1R},p_{2A},p_{3A}\ ; K) = f_1a_2a_4+f_2r_1a_4+f_4r_1r_2  \nonumber \\
\tilde{G}_{Ro} &=& \tilde{G}_R(p_{1A},p_{2A},p_{3R}\  ; K) = f_1a_2r_4+f_2r_1r_4+f_4a_1a_2 
\label{2.12}
\end{eqnarray}
The set (\ref{2.12}) of equations (and Fig.\ref{fi1}b) is, with Eq.(\ref{2.11}), identical to  Eqs.(66a,b,c) (and Fig.2)  of Ref.\cite{Ca1}, up to a term $r_1r_2r_4+a_1a_2a_4$ that integrates to zero in the $k_0$ integration.
Later on, the integrands associated with the advanced amplitudes will be needed. Here the $\epsilon$-flow is reversed along the tree's line,  i.e. 
$ r \leftrightarrow a $ in the integrand ,  keeping the $f_i$  {\it unchanged}. We will denote this operation with the sign $+$, i.e.
\begin{equation}
\tilde{G}_A \ = \ \tilde{G}_R^+ \ \ , \ \ \ \tilde{G}_{Ai} \ = \ \tilde{G}_{Ri}^+ \ \ , \ \ \ \tilde{G}_{Ao} \ = \ \tilde{G}_{Ro}^+
\label{2.12a}   
\end{equation}
just an in Eq.(\ref{2.2b}) one has $\tilde{\Sigma}_A=\tilde{\Sigma}_R^+$. Then,
\begin{eqnarray}
\tilde{G}_A &=& \tilde{G}_R(p_{1R},p_{2A},p_{3R} \ ; K) = f_1a_2a_4+f_2r_1r_4+f_4r_1a_2  \nonumber \\
\tilde{G}_{Ai} &=& \tilde{G}_R(p_{1A},p_{2R},p_{3R} \ ; K) = f_1r_2r_4+f_2a_1r_4+f_4a_1a_2  \nonumber \\
\tilde{G}_{Ao} &=& \tilde{G}_R(p_{1R},p_{2R},p_{3A}\  ; K) = f_1r_2a_4+f_2a_1a_4+f_4r_1r_2 
\label{2.13}
\end{eqnarray}
In Ref.\cite{Ca1}, the authors write three other amplitudes that are specific to the Keldysh basis. They obey the set ${\cal B}$ of rules in the case when only one external leg is of type $A$. There are three ways of cutting the triangle loop of Fig.\ref{fi1}b into two pieces, and two are allowed for that case. For example, if the flow is outgoing $P_2$, then the cut through $K_1, K_2$ is forbidden as the outgoing flow is not connected to an incoming flow. The rules lead to
\begin{eqnarray}
\tilde{G}_F &=& \tilde{G}_F(p_{1R},p_{2A},p_{3R} \ ; K) \nonumber \\
 &=& [f_2f_4-(r_2-a_2)(r_4-a_4)]r_1 \ +\ [f_1f_4-(r_1-a_1)(r_4-a_4)]a_2 \nonumber \\
 &=& f_2f_4r_1+f_1f_4a_2+r_4a_2a_1+a_4r_1r_2-r_1r_2r_4-a_1a_2a_4 \nonumber \\
\tilde{G}_{Fi} &=& \tilde{G}_F(p_{1A},p_{2R},p_{3R}\  ; K) \nonumber \\
 &=& [f_2f_1-(r_2-a_2)(r_1-a_1)]r_4 \ +\ [f_2f_4-(r_2-a_2)(r_4-a_4)]a_1 \nonumber \\
\tilde{G}_{Fo} &=& \tilde{G}_F(p_{1R},p_{2R},p_{3A}\  ; K) \nonumber \\
 &=& [f_2f_1-(r_2-a_2)(r_1-a_1)]a_4 \ +\ [f_1f_4-(r_1-a_1)(r_4-a_4)]r_2  
\label{2.14}
\end{eqnarray}
The set (\ref{2.14}) of equations is, with Eq.(\ref{2.11}), identical to Eqs.(66d,e,f) of Ref.\cite{Ca1} up to a term $r_1r_2r_4+a_1a_2a_4$ that integrates to zero.
\\ In the following, we concentrate on the thermal equilibrium case, and  we will come back to the non-equilibrium case at the end of 
Sec.\ref{se3.3}.

\subsubsection{The relations between Keldysh and Retarded-Advanced}
\label{se3.2.2}
If one considers the case of thermal equilibrium, one may use the relations (\ref{2.7})(\ref{2.8}) to transform the thermal distribution functions so that
\begin{eqnarray}
\tilde{G}_F=N_B(P_3)[f_2(r_4-a_4)-f_4(r_2-a_2)]r_1 \  \nonumber \\
 + \  N_F(P_1)[f_4(r_1-a_1)-f_1(r_4-a_4)]a_2
\label{2.15}
\end{eqnarray}
and similar ones for $ \tilde{G}_{Fi} , \ \tilde{G}_{Fo}$. The comparison with Eqs.(\ref{2.12})(\ref{2.13}) leads to
\begin{eqnarray}
G_F^{\mu} \ = \ N_B(P_3) \ ( \ G_A^{\mu} - G_{Ri}^{\mu}) \ + \ N_F(P_1) \ ( \ G_A^{\mu} - G_{Ro}^{\mu}) \nonumber \\
G_{Fi}^{\mu} \ = \ N_B(P_3) \ ( \ G_{Ai}^{\mu} - G_{R}^{\mu}) \ + \ N_F(P_2) \ ( \ G_{Ai}^{\mu} - G_{Ro}^{\mu}) \nonumber \\
G_{Fo}^{\mu} \ = \ N_F(P_1) \ ( \ G_{Ao}^{\mu} - G_{R}^{\mu}) \ + \ N_F(P_2) \ ( \ G_{Ao}^{\mu} - G_{Ri}^{\mu}) 
\label{2.16}
\end{eqnarray}
i.e. the Keldysh amplitudes $G_F, G_{Fi}, G_{Fo}$ are linear combinations of the retarded and advanced three-point amplitudes. For the scalar field case, the relations similar to the set (\ref{2.16}) were obtained in Ref.\cite{Ca3}, (for connected Green functions)  from the general dispersion relations derived from time ordering (see Eq.(33) in Ref.\cite{Ca3}). Alternatively, in perturbation theory, the set (\ref{2.16})   may be seen as a consequence of the linear matrix transformation that connects the Keldysh basis to the $R/A$ basis (see the comment after Eq.(\ref{2.10})). The set (\ref{2.16}) of equations is as general as the relation (\ref{2.10}) between $\Sigma_F$ and $\Sigma_R-\Sigma_A$.

\subsection{The interpretation of the Keldysh 3-point functions}
\label{se3.3}
\subsubsection {The analytical continuations of the IT amplitude}
\label{se3.3.1}
This section contains a reminder of useful properties \cite{Ev}.
\\ There is an unique imaginary time (IT) amplitude $G_{IT}^{\mu}(P_1, P_2, P_3)$. Here the components $p_1^0, p_2^0, p_3^0$ are imaginary, they take discrete values (interdistance $\pi T$) such that $p_1^0+p_2^0+p_3^0=0$. The analytical continuation towards real $p_i^0$ is made, keeping the constraint $p_1^0+p_2^0+p_3^0=0$ on both real and imaginary parts. One may continue $p_i^0$: i) either in the upper complex $p_i^0$ plane  to approach the real axis from above, i.e. $p_i^0 +i\epsilon$,   $ \epsilon>0$, ii) or in the lower plane to approach the real axis  from below, i.e. $p_i^0-i\epsilon$. The correspondance with the denomination type $R$ or type $A$ is $p_{iR}=(p_i^0+i\epsilon, {\bf p}_i)$,  $p_{iA}=(p_i^0-i\epsilon, {\bf p}_i)$  with $\epsilon>0$. 
The set of variables $p_{1R}, p_{2A}, p_{3A}$ refers to the analytical continuation from the imaginary values,  into  the domain 
${\rm Im} p_1^0 >0 , \  {\rm Im} p_2^0 <0 . \ {\rm Im} p_3^0<0 $,    towards the real values. The retarded amplitude $G^{\mu}(p_{1R}, p_{2A}, p_{3A})$ is thus one analytical continuation of the IT amplitude, and the $\epsilon$-flow along a tree keeps track of the momentum where the flow is incoming.

 A discontinuity in $p_1^0$, keeping $p_2^0$ and $p_3^0$ fixed, is the difference between two analytical continuations which only differ in the sign of ${\mathrm Im} p_1^0$, i.e. with the definition (\ref{2.6}) of the discontinuity
\begin{equation}
  {\mathrm Disc}_{p_1} G_R^{\mu}(p_1, p_{2R},p_{3A}) =G_R^{\mu}(p_{1R}, p_{2R}, p_{3A}) \ - \ G_R^{\mu}(p_{1A}, p_{2R}, p_{3A})
\label{2.18a}
\end{equation}
The possible analytical  continuations may be written $G^{\mu}(p_1^0+i\epsilon_1, p_2^0+i\epsilon_2, p_3^0+i\epsilon_3)$ where 
$\epsilon_i=\pm 1$, with the constraint $\epsilon_1 + \epsilon_2 + \epsilon_3 = 0$. As a consequence of this constraint, either one leg is of type $R$ and two legs are of type $A$ (the continuations are the retarded amplitudes), or two legs are of type $R$ and one leg is of type $A$ (they are the advanced amplitudes). To conclude, the retarded and advanced amplitudes are the six possible analytical continuations of the IT amplitude.
\\ 
It is worth emphasizing that, even though the three variables $p_1^0, p_2^0, p_3^0$ are linked by the relation $p_1^0+p_2^0+p_3^0=0$, one still has to consider discontinuities in each of these three variables, keeping the two other fixed  (if compatible with $\epsilon_1 + \epsilon_2 + \epsilon_3 = 0$). This case is the mathematical equivalent of the four-point elastic amplitude at $T=0$. Here the variables $s, t, u$ are linked by the relation $s+t+u=4m^2$ and one considers singularities and discontinuities of the diagrams in the $s$ ot $t$ or $u$ channel. At $T\not =0$, one may draw an $(s, t, u)$ plane for the variables  $p_1^0, p_2^0, p_3^0$ of the 3-point function.

\subsubsection{Keldysh functions and discontinuity in one variable}
\label{se3.3.2}
The relations (\ref{2.16}) involving the Keldysh amplitudes have an interesting interpretation, when one writes them in more precise terms.
For exemple, from (\ref{2.16}), with the precise denominations (\ref{2.12})(\ref{2.13})(\ref{2.14}), $G_{Fo}$ is
\begin{eqnarray}
G_{Fo}^{\mu}&=&G_F^{\mu}(p_{1R}, p_{2R}, p_{3A}) \nonumber \\
&=& \ N_F(P_1) \ [ G_R^{\mu}(p_{1R}, p_{2R}, p_{3A})  \ - \ G_R^{\mu}(p_{1A}, p_{2R}, p_{3A})] \nonumber \\
& &+ \  N_F(P_2) \ [ G_R^{\mu}(p_{1R}, p_{2R}, p_{3A}) \ - \ G_R^{\mu}(p_{1R}, p_{2A}, p_{3A})
\label{2.18}
\end{eqnarray} 
The factor that multiplies $N_F(P_1)$ is the discontinuity in the $p_1^0$ variable, keeping the $p_2^0$ and $p_3^0$ variables fixed to $p_{2R}$ and $p_{3A}$. Equation (\ref{2.18}) reads (with the definition (\ref{2.18a}) of the discontinuity)
\begin{eqnarray}
G_{Fo}^{\mu}&=& \ N_F(P_1)  \ {\mathrm Disc}_{p_1}G_R^{\mu}(p_1, p_{2R}, p_{3A}) \  \nonumber \\
& &+ \  N_F(P_2) \ {\mathrm Disc}_{p_2}G_R^{\mu}(p_{1R}, p_2, p_{3A})  \nonumber \\
&=&G_{Fo}^{\mu}(p_1,p_2,p_{3A})
\label{2.19}
\end{eqnarray}
$G_{Fo}$ is the sum of two discontinuities in one energy variable, while the energy variable $p_3^0$ is being kept fixed to $p_{3A}$. The last denomination  emphasizes this fact. Similarly, from Eq.(\ref{2.16}) with Eqs.(\ref{2.12})(\ref{2.13})(\ref{2.14}),
\begin{eqnarray}
G_{Fi}^{\mu} &=& \ N_B(P_3) \  {\mathrm Disc}_{p_3}G_R^{\mu}(p_{1A}, p_{2R}, p_3 ) \  \nonumber \\
  & &+ \  N_F(P_2) \  {\mathrm Disc}_{p_2}G_R^{\mu}(p_{1A}, p_2, p_{3R})   \nonumber \\
 &=& G_{Fi}^{\mu}(p_{1A},p_2,p_3)
\label{2.20a}
\end{eqnarray} 
\begin{eqnarray} 
G_{F}&=& \ N_B(P_3)  \ {\mathrm Disc}_{p_3}G_R^{\mu}(p_{1R}, p_{2A}, p_{3}) \  \nonumber \\
& &+ \  N_F(P_1) \  {\mathrm Disc}_{p_1}G_R^{\mu}(p_{1}, p_{2A}, p_{3R})  \nonumber \\
&=&G_{F}^{\mu}(p_1,p_{2A},p_{3})
\label{2.20b}
\end{eqnarray} 
One feature emerges from Eqs.(\ref{2.19})(\ref{2.20a})(\ref{2.20b}) for the amplitudes that are specific to the Keldysh basis. They are the sum of two discontinuities in one energy variable and the discontinuity in $p_i^0$ is weighted by $N(P_i)$. The feature is present in the 2-point function $\Sigma_F$ (see Eqs.(\ref{2.10})(\ref{2.6})) and it will carry over to the 4-point functions. 
Moreover, as it was reminded in Sec.\ref{se3.1},  to take a discontinuity of a diagram in a channel $P_i$ corresponds to put on-shell all particles  in one intermediate state of that channel, and the role of $N(P_i)$ is to transform the intermediate state's weight  from "emission $\mp$ absorption" into "emission $\pm$ absorption", i.e. the sign opposite to the one that appears in the retarded amplitudes.
\\ Let us stress the generality of this interpretation of the 3-point amplitudes $G_F, G_{Fi}, G_{Fo}$ as a sum of two discontinuities in one energy variable, keeping the two other energies fixed. It is as general as Eq.(\ref{2.16}).

One now considers the $G_F$ functions in the out-of-equilibrium case (here no analyticity argument can be used). What survives is the interpretation of intermediate states on-shell in a given channel of a diagram, together with the intermediate state's weight  "emission $\pm$ absorption"  (i.e. with a sign opposite to the one in the retarded functions $G_R$). It is true for the one-loop case, as obtained in Ref.\cite{Ca1} or from the set ${\cal B}$ of rules. One expect it to be general for $G_F, G_{Fi}, G_{Fo}$.

For the equilibrium case, the origin of the set ${\cal B}$ of rules for one-loop diagrams now appears. In order to compute in Eq.(\ref{2.18}) the discontinuity of $G_R$ in $p_1^0$, one considers both cases $p_{1R}$ and $p_{1A}$, keeping fixed $p_{2R}, p_{3A}$. One computes the one-loop diagram with the set ${\cal A}$ of rules for both cases. Their difference, weighted by $N(P_1)$,  is indeed the term with momenta  $K_1, K_2$ on-shell in Eq.(\ref{2.14}) for $G_{Fo}$, if one uses Eq.(\ref{2.8}) (see Sec.4 in Ref.\cite{FG1}). The sets ${\cal A}, {\cal B}$ of rules may be generalized to multiloop diagrams.

\subsection{The last 3-point vertex}
\label{se3.4}
In addition to the three retarded vertices $G_R, G_{Ri}. G_{Ro}$ and the three Keldysh vertices $G_F, G_{Fi}, G_{Fo}$, the Keldysh basis possesses a seventh vertex $G_E$, a necessity for the out-of-equilibrium case.

In thermal equilibrium, $G_E$ must be a linear combination of the six vertices of the $R/A$ basis, i.e. the retarded and advanced amplitudes. The relation was established in Ref.\cite{Ca3} for the scalar case, together with the ones for $G_F, G_{Fi}. G_{Fo}$, for connected functions. Translated to the QED vertex and to our notations, the relation reads
\begin{eqnarray}
G_E \ =\  \ N_F(P_2) N_B(P_3) \  ( \ G_{Ri} + G_{Ai} \ ) + N_F(P_1)N_B(P_3) \ ( \ G_R+G_A \ )  \nonumber \\
+N_F(P_1)N_F(P_2) \ ( \ G_{Ro}+G_{Ao} \ ) + G_{Ai} + G_A +G_{Ao}
\label{2.32}
\end{eqnarray}
where the functions such as $G_A=G_R^+$ have been defined in Eq.(\ref{2.12a}). For example, for the one-loop diagram of Fig.\ref{fi1}b, the integrand is obtained from the expressions for the functions $G_R, G_A$, i.e. Eqs.(\ref{2.12})(\ref{2.13}), and it has a term trilinear in the distribution functions
\begin{equation}
\tilde{G}_E = (r_1-a_1)(r_2-a_2)(r_3-a_3)N_1N_2N_3 + f_1r_2a_4+f_2a_1r_4+f_4a_2r_1
\end{equation}
where  $N_i=N(K_i)$, and where the following relations may be deduced from Eq.(\ref{2.7})
\begin{eqnarray}
\lefteqn{-1 \ $=$ \  N(P_1)N(P_2) + N(P_2)N(P_3) + N(P_1)N(P_3)}     \label{2.33} \\
- \ N_1N_2N_4  \ &=& \ N(P_1)N(P_2) N_1 + N(P_2)N(P_3) N_2 + N(P_1)N(P_3) N_4  \label{2.34} 
\end{eqnarray}
Other relations between the one-loop integrands are of interest
\begin{equation}
\tilde{G}_A-\tilde{G}_{Ri} =  f_2r_1(r_4-a_4)+f_4r_1(a_2-r_2) 
\end{equation}
so that 
\begin{eqnarray}
(\tilde{G}_A-\tilde{G}_{Ri}) - (\tilde{G}_{Ai}-\tilde{G}_{R}) &=& (N_2-N_4)(r_1-a_1)(r_2-a_2)(r_4-a_4)    \label{2.35}  \\
(\tilde{G}_{Ai}-\tilde{G}_{Ro}) - (\tilde{G}_{Ao}-\tilde{G}_{Ri}) &=& (N_1-N_2)(r_1-a_1)(r_2-a_2)(r_4-a_4)    \label{2.36}  \\
(\tilde{G}_{Ao}-\tilde{G}_{R}) - (\tilde{G}_{A}-\tilde{G}_{Ro}) &=& (N_4-N_1)(r_1-a_1)(r_2-a_2)(r_4-a_4)    \label{2.37}  
\end{eqnarray}  
With the more precise definitions in Eqs.(\ref{2.12})(\ref{2.13}), one has a general interpretation of these quantities. $G_A-G_{Ri} =G(p_{1R},p_{2A}, p_{3R})-G(p_{1R},p_{2A}, p_{3A})$ is the discontinuity in $p_3^0$, keeping the variables $p_{1R}^0$ and $ p_{2A}^0$ fixed. The result is that particles are on-shell in an intermediate state of the $p_3^0$ channel. Then, Eq.(\ref{2.35}) computes the discontinuity of this discontinuity 
\begin{eqnarray}
( G_A-G_{Ri} ) -( G_{Ai}-G_{R} ) = [ G(p_{1R},p_{2A}, p_{3R})-G(p_{1R},p_{2A}, p_{3A}) ] \nonumber \\
 -[ G(p_{1A},p_{2R}, p_{3R})-G(p_{1A},p_{2R}, p_{3A}) ]
\end{eqnarray} 
Indeed, in both terms, one computes $F(p_{3R})-F(p_{3A})$ (i.e. the diagram is cut in the $P_3$ channel), and one substracts the case when the $\epsilon$-flow goes from $P_2$ to $P_1$, from the case when it  flows from $P_1$ to $P_2$  (the cut lines are ignored by the flow). The result of that substraction again obeys Cutkosky rules (see after Eq.(\ref{2.6})), the particles are on-shell in an intermediate state of the $P_1 \rightarrow P_2$ channel. The result is, the diagram is now cut into three pieces (see Sec.5 and Fig.11 in Ref.\cite{FG1}). In short, the quantity 
$( G_A-G_{Ri} ) -( G_{Ai}-G_{R} )$ is a double discontinuity (in the sense of the Mandelstam double discontinuity in the $s, t, u$ plane). From the left-hand side of Eqs.(\ref{2.35})(\ref{2.36})(\ref{2.37}), one sees that the three double-discontinuities add up to zero, i.e. only two of them are linearly independent.  Any two of them can be made to appear in   $G_E$, as Eq.(\ref{2.32})  may be written
\begin{eqnarray} 
G_E = N_F(P_1)N_B(P_3) \ ( \ G_R + G_A- G_{Ri}-G_{Ai} \ )  \nonumber   \\
+ N_F(P_1)N_F(P_2) \ ( \ G_{Ro} + G_{Ao}- G_{Ri}-G_{Ai} \ ) +G_A+G_{Ao}-G_{Ri} 
\end{eqnarray}
On the right-hand side, in the first term $N_B(P_3)$ transforms the weight of the intermediate state in the $P_3$ channel, and  so does $N_F(P_2)$ in the $P_2$ channel in the second term. Then $N_F(P_1)$ should act on the sum of the terms so that the weight of a diagram's cut into three pieces does not depend on the choice in the cuts. 

To summarize, $G_E$ contains the double discontinuities of an arbitrary three-point diagram, i.e. all ways to cut the diagram into three pieces, where  an external leg is attached to each piece. 

\subsection{Ward identities between 2- and 3-point functions : IT versus Keldysh}
\label{se3.5}
In Ref.\cite{Ca3},  it is shown that three Ward identities relate the 3-point functions $G_R, G_{Ri}, G_{Ro}$ to the 2-point one $\Sigma_R$, and three identities relate $G_F,G_{Fi}, G_{Fo}$ to $\Sigma_F$. The relations were obtained from simple algebraic identities relating the one-loop integrands, once the approximation $P_i \ll K$ is made in the spin part of the integrand. This section will show that, in the equilibrium case,  those six Ward identities are analytical continuations of the unique, general, Imaginary Time (IT) Ward identity. Retarded functions $G$ will be related to retarded $\Sigma$, and  discontinuities will be related to discontinuities.

The usual way of writing the IT Ward identity is
\begin{equation}
K_{\mu} \ \Gamma^{\mu}(K,P) =\  e \ [ \ \Sigma(P) - \Sigma(P+K) \ ]
\label{2.21}
\end{equation}
where $P$ is the incoming fermion momentum and $K$ the incoming photon momentum. In our notation, Eq.(\ref{2.21}) reads
\begin{eqnarray}
P_{3\mu} \ G_{IT}^{\mu}(P_1, P_2, P_3) &=& e \ [ \ \Sigma(P_1) - \Sigma(P_1+P_3) \ ]  \label{2.22} \\
\Sigma(P_1+P_3) &=& \Sigma(-P_2)
\end{eqnarray}
\subsubsection{Identities involving retarded functions}
\label{se3.5.1}
The three retarded 3-point functions are obtained from the IT function via the analytic  continuation described in Sec.\ref{se3.3.1}. One just has to analytically continue both members of Eq.(\ref{2.22}). For example, the approach $p_{1A}, p_{2A}, p_{3R}$ towards the real plane is through the domain ${\mathrm Im} p_1^0  < 0 \ , \  {\mathrm Im} p_2^0  < 0 \ , \ {\mathrm Im} p_3^0  > 0 $. Ref.\cite{Ca2}'s authors use the momenta $P_1$ and $-P_2$, and one has to note that $p_{2A}$,  i.e. $p_2^0-i\epsilon, \ \epsilon>0$, corresponds to $(-p_2)_R$, i.e. $-p_2^0+i\epsilon$. The analytical continuations of Eq.(\ref{2.22}) give
\begin{eqnarray}
P_{3\mu} \ G_{Ro}^{\mu} \ &=& \  P_{3\mu} \ G_{R}^{\mu}(p_{1A}, p_{2A}, p_{3R}) \nonumber \\
 &=& \  -ie  \ [ \ \Sigma_A(P_1)-\Sigma_R(-P_2) \ ]  \label{2.23} \\ 
P_{3\mu} \ G_{R}^{\mu} \ &=& \  P_{3\mu} \ G_{R}^{\mu}(p_{1A}, p_{2R}, p_{3A}) \nonumber \\
 &=& \  -ie  \ [ \ \Sigma_A(P_1)-\Sigma_A(-P_2) \ ]  \label{2.24} \\ 
P_{3\mu} \ G_{Ri}^{\mu} \ &=& \  P_{3\mu} \ G_{R}^{\mu}(p_{1R}, p_{2A}, p_{3A}) \nonumber \\
 &=& \  -ie  \ [ \ \Sigma_R(P_1)-\Sigma_R(-P_2) \ ]  \label{2.25} 
\end{eqnarray}   
(the change $e \rightarrow -ie$ comes from the vertex definitions in Euclidean and Minkowski spaces, see Sec.A.4 in Ref.\cite{MLB}). This set of relations may be compared to the one found from algebraic identities in Ref.\cite{Ca2}. One should be aware of the different notations in Ref.\cite{Ca1} and Ref.\cite{Ca2}. To go from Ref.\cite{Ca1} (i.e. our notations) to Ref.\cite{Ca2}, the correspondence in external momenta  is $P_1=P, P_2=-Q, P_3 =K$ and the substitution in the integrands is $1\rightarrow s \ , \  4\rightarrow 1 \ , \ 2\rightarrow 3 $ together with $r\rightarrow a \ , \ a\rightarrow r$ (compare Fig.1 of Ref.\cite{Ca1} and Fig.2 of Ref.\cite{Ca2}). As a result, the correspondence between the names for the amplitudes written in Eq.(66) of Ref.\cite{Ca1} and Eqs.(33)(34) of Ref.\cite{Ca2}  is 
\begin{eqnarray} 
G_{Ro}=\Gamma_R \ , \ G_R=\Gamma_{R_i} \ , \ G_{Ri}=\Gamma_{Ro}  \nonumber \\ 
G_{Fo}=\Gamma_F \ , \ G_F=\Gamma_{F_i} \ , \ G_{Fi}=\Gamma_{Fo} \label{2.26} 
\end{eqnarray}
With Relations (\ref{2.26}), one can check that the identities  (\ref{2.23})(\ref{2.24})(\ref{2.25}) are indeed the first three equations of the set (38) in Ref.\cite{Ca2}.
\subsubsection{Identities involving Keldysh functions}
\label{se3.5.2} 
The other identities involve the functions $G_F$ which are a sum of two discontinuities of $G_R$ functions.
\\ i) $G_{Fo}$ is known from Eq.(\ref{2.18}), so that
\begin{eqnarray}
P_{3\mu}G_{Fo}^{\mu}= N_F(P_1) \ P_{3\mu} \ [ \ G_{R}^{\mu}(p_{1R}, p_{2R}, p_{3A}) - G_{R}^{\mu}(p_{1A}, p_{2R}, p_{3A}) \ ] \nonumber \\ 
 + N_F(P_2) \ P_{3\mu} \ [ \ G_{R}^{\mu}(p_{1R}, p_{2R}, p_{3A}) - G_{R}^{\mu}(p_{1R}, p_{2A}, p_{3A}) \ ]
\label{2.27} 
\end{eqnarray}
$P_{3\mu}G_{R}^{\mu}(p_{1A}, p_{2R}, p_{3A})$ is written in Eq.(\ref{2.24}). When one performs the substraction in order to get the discontinuity in $p_1^0$, the term $\Sigma(-P_2)$ disappears, as $p_{2R}, p_{3A}$ are being kept fixed. A similar fact occurs for the discontinuity in $p_2^0$ with  $P_{3\mu}G_{R}^{\mu}(p_{1R}, p_{2A}, p_{3A})$ as in Eq.(\ref{2.25}). Then Eq.(\ref{2.27}) becomes
\begin{eqnarray}
P_{3\mu}G_{Fo}^{\mu}&=&-ie \ [ \ N_F(P_1) \ (\Sigma_R(P_1)-\Sigma_A(P_1) ) \nonumber \\
&\ &+N_F(P_2) \ (-\Sigma_A(-P_2)+\Sigma_R(-P_2) ) \ ] \nonumber \\
P_{3\mu}G_{Fo}^{\mu} &=& -ie \ [ \ \Sigma_F(P_1) - \Sigma_F(-P_2) \ ]
\label{2.28}
\end{eqnarray}
with $\Sigma_F(P)$ from Eq.(\ref{2.10}) and $N_F(P_2)= - N_F(-P_2)$.
\\ 
ii) For the amplitude $G_F^{\mu}= G_{F}^{\mu}(p_{1R}, p_{2A}, p_{3R})$, $P_2$ is being kept fixed to $p_{2A}$, while discontinuities are taken in $p_1^0$ and $p_3^0$. Then the term in $\Sigma(-P_2)$ disappears from $P_{3\mu}G_{Ro}^{\mu}$ in the substraction that results from taking the discontinuity in $p_1^0$. Also the discontinuity in $p_3^0$ is zero, as $P_3$ is not a variable which  appears on the right-hand side of $P_{3\mu}G_{R}^{\mu}$ or $P_{3\mu}G_{Ri}^{\mu}$ or $P_{3\mu}G_{Ro}^{\mu}$ in Eqs.(\ref{2.23})(\ref{2.24})(\ref{2.25}). Then,
\begin{eqnarray}
P_{3\mu}G_{F}^{\mu} &=& N_F(P_1) \ P_{3\mu} \ [ \ G_{R}^{\mu}(p_{1R}, p_{2A}, p_{3R}) - G_{R}^{\mu}(p_{1A}, p_{2A}, p_{3R}) \ ]
\nonumber \\
&=& -ie \ N_F(P_1) \  ( \Sigma_R(P_1)-\Sigma_A(P_1) ) \nonumber \\
P_{3\mu}G_{F}^{\mu} &=&-ie \ \Sigma_F(P_1)
\label{2.29}
 \end{eqnarray}
 iii)
Similarly for $G_{Fi}^{\mu}=G_{F}^{\mu}(p_{1A}, p_{2R}, p_{3R})$, $p_{1A}$ is being kept fixed, while discontinuities are taken in $p_2^0$ and $p_3^0$. The term $\Sigma(P_1)$ disappears from $P_{3\mu}G_{Ro}^{\mu}$ and the discontinuity in $p_3^0$ is zero.
\begin{eqnarray}
P_{3\mu}G_{Fi}^{\mu} &=& -ie \ N_F(P_2) \  ( -\Sigma_A(-P_2)-\Sigma_R(-P_2) ) \nonumber \\ 
P_{3\mu}G_{Fi}^{\mu}&=& ie \ \Sigma_F(-P_2)
\label{2.30}
\end{eqnarray}
Eqs.(\ref{2.28})(\ref{2.29})(\ref{2.30}) may be seen to agree with the last three Equations (38) of Ref.\cite{Ca2}, if one uses the correspondence (\ref{2.26}). Two general observations will be used later on: i) in the contraction $P_{3\mu}G_R^{\mu}$, $P_3$ disappears as an independent variable, $P_3$ only appears in  $P_1+P_3=-P_2$ and in  $P_2+P_3=-P_1$. A consequence is that the discontinuity in $p_3^0$ is zero in the contraction of $P_3^{\mu}$ with the $G_F$ functions. ii) When one variable is being kept fixed in taking a discontinuity, the terms that depend only on that fixed variable, disappear from the discontinuity. \\
To summarize, the six Ward identities are just rewritings of the IT Ward identity, in the case of thermal equilibrium.

\section{Four-point amplitudes}
\label{se4}
One follows a track similar to the three-point amplitudes' case.
\subsection{One-loop integrands}
\label{se4.1}
\begin{figure}
\includegraphics*[width=10cm]{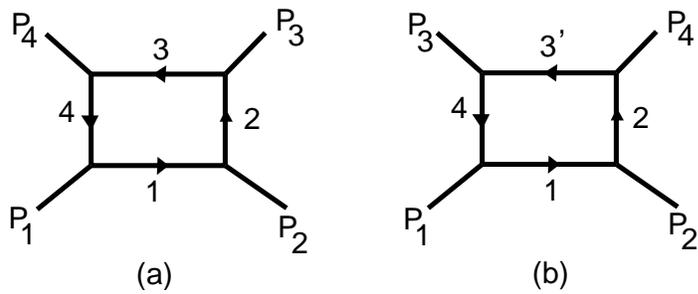}
\caption{\label{fi2} One-loop diagrams for the 2-photon-2-fermion QED vertex . $P_1$ is the incoming fermion momentum, $P_3, P_4$ are  incoming photon momenta.  1, 2, 3, 4 refer to momenta in the loop, 1 is a photon, the other ones are fermions.}
\end{figure}

The one-loop QED amplitude involves two fermion external momenta $P_1, P_2$ and two photon external momenta $P_3, P_4$. The two one-loop diagrams to be added are drawn on Fig.\ref{fi2}. Most of the discussion will concentrate on diagram (a). With the approximation in the factor that involves the $\gamma$ matrices, 
\begin{equation}
G^{\mu \nu} = 4 e^4 \int {d_4K\over{(2\pi)^4}} \ K^{\mu} K^{\nu} \ {\gamma}.K \ \ \tilde{G}
\label{4.1}
\end{equation}
The set ${\cal A}$ of rules of Sec.\ref{se2} gives the retarded amplitudes' integrands. There are four ways of cutting once the loop momentum, i.e. the integrand is the sum of four terms
\begin{eqnarray} 
\tilde{G}_{R1}^{(a)} &=& \tilde{G}_{R}^{(a)} (p_{1R}, p_{2A}, p_{3A}, p_{4A} \ ; K) \nonumber \\
&=&f_1a_2a_3a_4+f_2r_1a_3a_4+f_3r_1r_2a_4+f_4r_1r_2r_3  \nonumber \\
\tilde{G}_{R2}^{(a)} &=& \tilde{G}_{R}^{(a)} (p_{1A}, p_{2R}, p_{3A}, p_{4A} \ ; K) \nonumber \\
&=&f_1r_2r_3r_4+f_2a_1a_3a_4+f_3a_1r_2a_4+f_4a_1r_2r_3  \nonumber \\
\tilde{G}_{R3}^{(a)} &=& \tilde{G}_{R}^{(a)} (p_{1A}, p_{2A}, p_{3R}, p_{4A} \ ; K) \nonumber \\
&=&f_1a_2r_3r_4+f_2r_1r_3r_4+f_3a_1a_2a_4+f_4a_1a_2r_3  \nonumber \\
\tilde{G}_{R4}^{(a)} &=& \tilde{G}_{R}^{(a)} (p_{1A}, p_{2A}, p_{3A}, p_{4R}\ ; K) \nonumber \\
&=&f_1a_2a_3r_4+f_2r_1a_3r_4+f_3r_1r_2r_4+f_4a_1a_2a_3  
\label{4.2}
\end{eqnarray}  
Together with Eq.(\ref{4.1}), the set (\ref{4.2}) of equations is identical to Eqs.(34a.b.c.d) and Eq.(36) of Ref.\cite{Ca1}, up to a term $a_1a_2a_3a_4+r_1r_2r_3r_4$ that integrates to zero in the $k^0$ integration. 

In Ref.\cite{Ca1}, the authors consider three other Keldysh 4-point amplitudes, which are called $G_{rara}^{\mu \nu}$,  $G_{rraa}^{\mu \nu}$,  $G_{raar}^{\mu \nu}$.  They are given by the set ${\cal B}$ of rules. There are two momenta with outgoing $\epsilon$-flow, and there are six ways of cutting the loop diagram (a)  of Fig.\ref{fi2} into two trees. The number of allowed ways (i.e. outgoing flow connected to an incoming one) are easily found. If the two outgoing-flow momenta are adjacent ones, there are three ways, and the integrand is the sum of three terms; if they are not adjacent ones, there are four ways.
\begin{eqnarray} 
\tilde{G}_{raar}^{(a)}&=&\tilde{G}_F^{(a)}(p_{1R}, p_{2A}, p_{3A}, p_{4R} \ ; K) \nonumber \\
 &=&[f_1f_4-(r_1-a_1)(r_4-a_4)]a_2a_3 + [f_3f_4-(r_3-a_3)(r_4-a_4)]r_1r_2 \nonumber \\
 & \ &+ [f_2f_4-(r_2-a_2)(r_4-a_4)]r_1a_3 
\label{4.4}
\end{eqnarray}
\begin{eqnarray}
\tilde{G}_{rraa}^{(a)}&=&\tilde{G}_F^{(a)}(p_{1R}, p_{2R}, p_{3A}, p_{4A} \ ; K) \nonumber \\
 &=&[f_1f_2-(r_1-a_1)(r_2-a_2)]a_3a_4 + [f_1f_4-(r_1-a_1)(r_4-a_4)]r_2r_3 \nonumber \\
&\ &+ [f_1f_3-(r_1-a_1)(r_3-a_3)]r_2a_4 
\label{4.5}
\end{eqnarray}
\begin{eqnarray}
\tilde{G}_{rara}^{(a)}&=&\tilde{G}_F^{(a)}(p_{1R}, p_{2A}, p_{3R}, p_{4A} \ ; K) \nonumber \\
 &=&[f_1f_3-(r_1-a_1)(r_3-a_3)]a_2a_4 + [f_2f_4-(r_2-a_2)(r_4-a_4)]r_1r_3 + \nonumber \\
& \ & [f_1f_4-(r_1-a_1)(r_4-a_4)]r_3a_2  + [f_2f_3-(r_2-a_2)(r_3-a_3)]r_1a_4
\label{4.6}
\end{eqnarray}
Together with Eq.(\ref{4.1}), Eqs.(\ref{4.4})(\ref{4.5})(\ref{4.6}) reduce respectively to Eqs.(34g)(34f)(34e) and Eq.(36) of Ref.\cite{Ca1}.

\subsection{The interpretation in terms of discontinuity}
\label{se4.2} 
For the thermal equilibrium case, the statistical weights of the cut lines  are transformed with the help of Eq.(\ref{2.8}) so that a weight involving the external momenta appears. With $P_1=K_1-K_4,\  P_4=K_4-K_3, \ P_1+P_2=K_2-K_4$, Eq.(\ref{4.4}) turns into
\begin{eqnarray}
\tilde{G}_{raar}^{(a)}&=&\tilde{G}_F^{(a)}(p_{1R}, p_{2A}, p_{3A}, p_{4R} \ ; K) \nonumber \\
&=& - N_F(P_1) \ [ \ f_1(r_4-a_4)-f_4(r_1-a_1)\  ]a_2a_3 \nonumber \\
& \ &-N_B(P_4) \ [ \ f_4(r_3-a_3)-f_3(r_4-a_4) \ ] r_1r_2 \nonumber \\
& \ & -N_B(P_1+P_2) \ [ f_2(r_4-a_4)-f_4(r_2-a_2) \ ] r_1a_3
\label{4.7}
\end{eqnarray} 
The interpretation of Eq.(\ref{4.7}) is similar to the one encountered in Sec.\ref{se3.3.2} for the 3-point Keldysh amplitudes $G_F$. $ G_F^{\mu \nu}(p_{1R}, p_{2A}, p_{3A}. p_{4R})$ is the sum of three discontinuities: a discontinuity in the $p_1^0$ variable (internal lines $K_1 , K_4$ on shell) weighted by $N_F(P_1)$, a discontinuity in the $p_4^0$ variable (internal lines $K_4, K_3$ on shell) weighted by $N_B(P_4)$, and a discontinuity in the $p_1^0+p_2^0$ variable (internal lines $K_2, K_4$ on shell) weighted by $N_B(P_1+P_2)$. In addition to the discontinuities in the external energy variables $p_i^0$ of type $R$ (keeping all the other variables fixed), one encounters a discontinuity in the subenergy $p_1^0+p_2^0=-(p_3^0+p_4^0)$. The occurence of discontinuities in subenergies will be fully discussed in Sec.\ref{se4.3.1}. Note that the definition of the discontinuity in Eq.(\ref{2.6}) and the parity property of $N(P_1)$ are such that
\begin{equation}
N(P_1+P_2)  \ {\mathrm Disc}_{p_1+p_2} G \ = \  N(P_3+P_4)  \ {\mathrm Disc}_{p_3+p_4} G
\label{4.8}
\end{equation}
Similarly $ \tilde{G}_{rraa}^{(a)}$ is the sum of three discontinuities and $\tilde{G}_{rara}^{(a)}$ is the sum of four discontinuities
\begin{eqnarray}
\tilde{G}_{rara}^{(a)}&=&\tilde{G}_F^{(a)}(p_{1R}, p_{2A}, p_{3R}, p_{4A} \ ; K) \nonumber \\
&=& - N_F(P_1) \ [ \ f_1(r_4-a_4)-f_4(r_1-a_1)\  ]a_2r_3 \nonumber \\
& \ &-N_B(P_3) \ [ \ f_3(r_2-a_2)-f_(r_3-a_3) \ ] r_1a_4 \nonumber \\
& \ & -N_B(P_1+P_2) \ [ f_2(r_4-a_4)-f_4(r_2-a_2) \ ] r_1r_3 \nonumber \\
& \ &-N_F(P_2+P_3) \ [ f_3(r_1-a_1)-f_3(r_1-a_1) \ ] a_2a_4
\label{4.9}
\end{eqnarray}  
If one now considers diagram (b) of Fig.\ref{fi2}, the external legs $P_3$ and $P_4$ are interchanged, so that $P_1$ and $P_3$ are now adjacent legs, while $P_1$ and $P_4$ are not adjacent ones. $G_{raar}^{(b)}$ is a sum of four discontinuities and $G_{rara}^{(b)}$  is a sum of three discontinuities. If one now adds diagrams (a) and (b), $G_{raar}^{(a)+(b)} =G_F^{(a)+(b)}(p_{1R}, p_{2A}, p_{3A}, p_{4R})$ has discontinuities in $P_1, P_4$ and in $ P_1+P_2, P_1+P_3 $, while $G_{rara}^{(a)+(b)} =G_F^{(a)+(b)}(p_{1R}, p_{2A}, p_{3R}, p_{4A})$ has discontinuities in $P_1 , P_3 , P_1+P_2,P_1+P_4$. The rules is: there are discontinuities in the two external legs of type $R$, $p_{iR},p_{jR}$ and discontinuities in subenergies in the two channels "transverse" to $P_i+P_j$. 
\\
To conclude, the three 4-point Keldysh amplitudes selected by the authors have the same features as the 3-point Keldysh amplitudes $G_F^{\mu}$. In thermal equilibrium:  i) they are a sum of discontinuities in external energyies of type $R$ and in subenergies ii) the discontinuity's weight involves a thermal factor of the energy variable. 
The generalization to the non-equilibrium case is easy: i) "discontinuity" is equivalent to "cut diagram", i.e. intermediate states on-shell in the appropriate energy or subenergy variable, ii) the weight of the intermediate state is the combination "emission $\pm$ absorption" opposite to the combination that appears in the retarded amplitudes. \\

\noindent {\bf  The hard-loop-soft-external  case} \\  
One  now focuses, in Ref.\cite{Ca1},  on the transition from their Eq.(34e) for  $G_{rara}$, to their Eqs.(46) (47a,b,c,d), where the coefficients $c_{\beta}, d_{\beta}$ show up, with values from their Eqs.(48)(49). In this hard-loop approximation, the external momenta  are neglected, in front of the loop momentum, in the statistical weights $[\ f_if_j-(r_i-a_i)(r_j-a_j) \ ] = [N(K_i)N(K_j) -1](r_i-a_i)(r_j-a_j )$. One writes $N(K_i)=N(K)$ for all $i$, with the result, from Eq.(\ref{1.5}),
\begin{eqnarray}
1-N_B(K)N_F(K) &=& 0 \nonumber \\
1-N_F(K)N_F(K) &=& {1\over{ \cosh^2{\beta\over2}k_0}}  \nonumber \\
\int_0^{\infty}dk \ k \ {1\over{ \cosh^2{\beta\over2}k_0}}&=& 4T^2 \ \ln 2
\end{eqnarray}
In the limit $K_i \rightarrow K$, the characteristic properties of the thermal weights, Eq.(\ref{2.7}), write
\begin{eqnarray}
N_F(K_1-K_2)[N_B(K_1)-N_F(K_2)] &\approx& {\beta\over2}(k_1^0-k_2^0)[N_B(K)-N_F(K)] \rightarrow 0  \nonumber \\
 N_B(K_3-K_2)[N_F(K_3)-N_F(K_2)] &\approx& {1\over{\beta\over2}(k_3^0-k_2^0)}[N_F(k_3^0)-N_F(k_2^0)] \nonumber \\
&\rightarrow& {2\over \beta}{dN_F\over{dk^0}} ={1\over{ \cosh^2{\beta\over2}k_0}}
\end{eqnarray}
When the amplitude $G_{rara}^{(a)}$ is written as in Eq.(\ref{4.9}), the weights involve soft external momenta. One sees that, in the hard loop approximation, the discontinuities in the fermionic channels are neglected. In the bosonic channels, a Taylor expansion has to be made in the statistical weight of the amplitude that multiplies the soft bosonic weight.

For consistency with the Ward identities (see the Appendix), similar approximations have to be made in the fermionic and bosonic channels of the Keldysh 2-point and 3-point functions i.e.   
\begin{equation}
\Sigma_F(P) \ =\  0  \ \ , \ \ \  \ \ G_F^{\mu}(p_{1R},p_{2R},p_{3A}) \ = \ 0
\end{equation}
If the two other $G_F^{\mu}$ functions are written under the general form (\ref{2.16}), a Taylor expansion has to be made in the statistical weights which appear inside the functions ($G_A$ and $G_{Ri}$), or ($G_{Ai}$ and $G_R$). The result involves a term which is usually neglected in the hard-loop approximation for the retarded amplitudes, hence the appearance of the overall $\ln 2 $ factor.

To conclude, the singularity in the soft bosonic weight brings to the forefront a term which is absent from the retarded amplitudes' HTL form. (For a hard-out-of-equilibrium situation with $\Sigma_F=0$, see ref.\cite{Ba})      

\subsection{Relating Keldysh to IT  via analytical continuations}
\label{se4.3}
In thermal equilibrium, the 4-point Keldysh amplitudes are linear combinations of analytic continuations of the unique Imaginary Time amplitude. The relation will be found from the inspection of the one-loop integrands, as it was the case for the relations (\ref{2.16})  found for the 3-point functions.

\subsubsection{The analytical continuations of a 4-point IT amplitude and the $R/A$ basis}
\label{se4.3.1}
i) {\bf the retarded or advanced amplitudes} \\
The retarded amplitudes are simple analytical continuations of the IT amplitudes. Indeed, the prescription on the external momenta $p_j^0+i\epsilon_j$ fixes the prescription on all subenergies $p_j^0+p_k^0$. For example for $G_R(p_{1R}, p_{2A}, p_{3A}. p_{4A})$, one has $\epsilon_1>0, \ \epsilon_2<0, \ \epsilon_3<0, \ \epsilon_4<0$, the imaginary part of $p_3^0+p_4^0$ is $\epsilon_3+\epsilon_4<0$, that of 
$p_1^0+p_2^0$ is $\epsilon_1+\epsilon_2=-(\epsilon_3+\epsilon_4) \ >0$. The advanced amplitudes correspond to the opposite case, i.e. one $\epsilon_i$ is negative, the other ones are positive. In thermal equilibrium, the advanced and retarded amplitudes are related (see Eq.(\ref{2.12a}))
\begin{equation}
 G(p_{1A}, p_{2R}, p_{3R}, p_{4R})=G_R^+(p_{1R}, p_{2A}, p_{3A}, p_{4A})
\label{5.1a}
\end{equation} 

\noindent ii) {\bf the other analytical continuations}\\
They correspond to the case when two $\epsilon$ are positive, and two are negative. There, the sign of two subenergies are not constrained by $\epsilon_1+\epsilon_2+\epsilon_3+\epsilon_4=0$  \cite{Ev}. For example, for $\epsilon_1<0, \ \epsilon_2>0, \ \epsilon_3>0, \ \epsilon_4<0$ the signs of $p_1^0+p_2^0$ and of $p_1^0+p_3^0$  may be $>0$ or $<0$, as they depend on the relative magnitude of the $\epsilon_i$ . However, the diagrams have branch cuts in those subenergies. (The change in the relative magnitude of the $\epsilon_i$  allows one to compute the discontinuity across the cut, keeping the signs of the $\epsilon_i$ fixed; those discontinuities are the one encountered in Sec.\ref{se4.2}). 

In an analytical continuation, it is necessary to know which way one is approching the real axis in those variables,  i.e. $(p_1+p_2)_R$ or $(p_1+p_2)_A$ and  $(p_1+p_3)_R$ or $(p_1+p_3)_A$. In thermal equilibrium, a precise answer exists in the $R/A$ formalism  for any $n$-point function; a weight is given to each possible sign for the subenergies \cite{FG1}. The rules will be stated  for the 4-point case, when the amplitude is written as in Eq.(\ref{5.1b})

The $R/A$ amplitudes obey the set $\cal A$ of rules of Sec.\ref{se2}. The one-loop diagram of Fig.\ref{fi2}a is written as a sum of 4 trees: i) with $K_1$ (or $K_3$) on shell, the tree depends on $P_1+P_4$, ii) with $K_4$ (or $K_2$) on shell, the tree depends on $P_1+P_2$ ; there is no dependence on $P_1+P_3$. The diagram of Fig.\ref{fi2}b is obtained via the substitution $P_3 \leftrightarrow P_4$ and depends on $P_1+P_3$ and $P_1+P_2$. The sum of the two diagrams may be written, in the IT formalism
\begin{equation}
G_{IT}^{(a)+(b)}(P_1,P_2,P_3,P_4)= F_1(P_1+P_2)+F_2(P_1+P_4)+F_3(P_1+P_3)
\label{5.1b}
\end{equation} 
where the $F_l$ also depend on the external momenta $P_i$.  Then, the amplitude in the $R/A$ formalism may be written, for example for the case $(p_{1A},p_{2R},p_{3R},p_{4A})$
\begin{eqnarray}
\lefteqn{{\mathcal N}(p_1,p_4) \ G_{R/A}^{(a)+(b)}(p_{1A},p_{2R},p_{3R},p_{4A})}  \nonumber  \\
&=&\ {\mathcal N}(p_1,p_4) \ F_2((p_1+p_4)_A)  \nonumber  \\
 &\ &+{\mathcal N}(p_1+p_2,p_4) \  F_1((p_1+p_2)_A) \ + \  {\mathcal N}(p_1,p_4+p_3) \  F_1((p_1+p_2)_R) \nonumber \\
&\ &+ {\mathcal N}(p_1+p_3,p_4) \  F_3((p_1+p_3)_A) \ + \  {\mathcal N}(p_1,p_2+p_3) \  F_3((p_1+p_3)_R)
\label{5.2}
\end{eqnarray} 
where
\begin{equation}
{\mathcal N}(p_i,p_j)\ = \ N(P_i) +N(P_j)
\label{5.3}
\end{equation}    
with $N(P_i)=-N(-P_i)$ as in Eq.(\ref{1.6}) (with the appropriate subscript boson/fermion), so that
\begin{equation}
{\mathcal N}(p_1,p_4+p_3) +{\mathcal N}(p_1+p_2,p_4) = {\mathcal N}(p_1,p_4)
\label{5.4}
\end{equation}
(Note that, in the weight ${\mathcal N}(p_i+p_j, p_k)$, there enter the subenergy $p_i+p_j$ which is of type $A$ in $F_l$ and one energy of type $A$ such that $ i\not = j \not = k$ ) The prescription (\ref{5.2}) satisfies the  relation \cite{vW, FG1}
\begin{equation}
G_{R/A}(p_{1R},p_{2A},p_{3A},p_{4R}) \ = \ G_{R/A}^+(p_{1A},p_{2R},p_{3R},p_{4A})
\label{5.6}
\end{equation}   
(The proof requires some algebra, together with two properties of ${\mathcal N}(p_i,p_j)$, see Eqs.(2.7) and (3.15) in \cite{FG1}).

\subsubsection{The one-loop case in the $R/A$ basis}
\label{se4.3.2}
The prescription (\ref{5.2}) complements the set $\cal A$ of rules for the $\epsilon$-flow along the tree. Now, two legs have entering flow, and it gives a rule for the propagator where the $\epsilon$-flow may be of either direction. The weight $\mathcal N$ arises from the tree's vertex that has two outgoing $\epsilon$-flow. For the diagram (a) of Fig.\ref{fi2}, the rules lead to
\begin{eqnarray}  
\lefteqn{ \tilde{G}_{R/A}^{(a)}(p_{1A},p_{2R},p_{3R},p_{4A}\ ;K) \  {\mathcal N}(p_1,p_4)} \nonumber \\
&=&  {\mathcal N}(p_1,p_4) \ [\ f_1r_2r_3r_4 + f_3a_1a_2a_4 \ ] \nonumber \\
&\ & + f_2r_3a_1 \ [ \ {\mathcal N}(p_1,p_4+p_3)a_4 + {\mathcal N}(p_1+p_2,p_4)r_4 \ ] \nonumber  \\
&\ & + f_4r_3a_1 \ [ \  {\mathcal N}(p_1,p_4+p_3)r_2 + {\mathcal N}(p_1+p_2,p_4)a_2 \ ]
\label{5.7}
\end{eqnarray}    
With Eq.(\ref{5.3}), Eq.(\ref{5.7}) may be written
\begin{eqnarray}
\lefteqn{\tilde{G}_{R/A}^{(a)}(p_{1A},p_{2R},p_{3R},p_{4A}\ ;K) \ [ \ N_F(P_1)+N_B(P_4) \ ]} \nonumber \\
&=& N_B(P_1+P_2) \ [ f_2r_3a_1(r_4-a_4) - f_4r_1a_3(r_2-a_2) ] \nonumber \\
&\ & + N_F(P_1) \ [ f_1r_2r_3r_4+f_3a_1a_2a_4+f_2a_1r_3a_4+f_4a_1r_2r_3 ] \nonumber \\
&\ & + N_B(P_4) \ [ f_1r_2r_3r_4+f_3a_1a_2a_4+f_2a_1r_3r_4+f_4a_1a_2r_3 ]
\label{5.8}
\end{eqnarray}
Another case is
\begin{eqnarray}
\lefteqn{\tilde{G}_{R/A}^{(a)}(p_{1A},p_{2R},p_{3A},p_{4R}\ ;K) \  {\mathcal N}(p_1,p_3) } \nonumber \\
& =& f_1r_2r_4 \ [ \ {\mathcal N}(p_1,p_2+p_3)a_3 + {\mathcal N}(p_1+p_4,p_3)r_3 \ ] \nonumber  \\
&\ & + f_3r_2r_4 \ [ \  {\mathcal N}(p_1,p_2+p_3)r_1 + {\mathcal N}(p_1+p_4,p_3)a_1 \ ] \nonumber \\
&\ & +f_2a_3a_1 \ [ \ {\mathcal N}(p_1,p_4+p_3)a_4 + {\mathcal N}(p_1+p_2,p_3)r_4 \ ] \nonumber  \\
&\ & + f_4a_3a_1 \ [ \  {\mathcal N}(p_1,p_4+p_3)r_2 + {\mathcal N}(p_1+p_2,p_3)a_2 \ ]
\label{5.9}
\end{eqnarray}  
Eq.(\ref{5.9}) may be transformed into a form similar to Eq.(\ref{5.8}), where there appear four weights: $N(P_1+P_2), N(P_1+P_4), N(P_1), N(P_3)$.

\subsubsection{The relation between Keldysh and $R/A$ functions}
\label{se4.3.3}   
With $G_F$ from Eq.(\ref{4.4}), $G_{Ri}$ from Eq.(\ref{4.2}), it is easy algebra to obtain the relation
\begin{eqnarray}
\lefteqn{G_F(p_{1R},p_{2A},p_{3A},p_{4R}) } \nonumber \\
&+&N_F(P_1) \ G_R(p_{1A},p_{2A},p_{3A},p_{4R}) +  N_B(P_4) \ G_R(p_{1R},p_{2A},p_{3A},p_{4A}) \nonumber \\
&=& [ \ N_F(P_1)+N_B(P_4) \ ] \ G_{R/A}^+(p_{1A},p_{2R},p_{3R},p_{4A})
\label{5.10}
\end{eqnarray}
where $G_{R/A}^+$ is obtained from Eq.(\ref{5.8}) by means of the substitution $ r \leftrightarrow a$ (it obeys Eq.(\ref{5.6})). Relation (\ref{5.10}) expresses the Keldysh amplitude $G_F$ as a linear combination of three analytical continuations of the IT amplitude. This relation may be understood in terms of discontinuities. In Sec.\ref{se4.2}, $ G_F^{(a)}(p_{1R},p_{2A},p_{3A},p_{4R})$ was shown to be the sum of three discontinuities: in $p_1^0$, in $p_4^0$, and in $p_1^0+p_2^0$. The addition to $G_F$ of the two functions $G_R$ suppresses the discontinuities in $p_1^0$ and $p_4^0$, leaving the discontinuity in $p_1^0+p_2^0$, which is explicit in form (\ref{5.8}) for $G_{R/A}^{(a)}$. The relation (\ref{5.10}) has been obtained from the inspection of the integrands associated with diagram (a). The relation is quite general, a consequence of the existence of a Bogoliubov transformation between the Keldysh and $R/A$ basis. The general $ G_F (p_{1R},p_{2A},p_{3A},p_{4R})$ is a sum of four discontinuities: in $p_1^0$, in $p_4^0$, in $p_1^0+p_2^0$, and in $p_1^0+p_3^0$. The addition of the two functions $G_R$ suppresses the discontinuities in $p_1^0$ and $p_4^0$ (Note that in Eq.(\ref{5.10}), each retarded function $G_R$ differs from $G_F$ only in one variable). The general $G_{R/A}(p_{1R},p_{2A},p_{3A},p_{4R})$ is one specific analytical continuation of the IT amplitude, and it has discontinuities in $p_1^0+p_2^0$, and in $p_1^0+p_3^0$, with appropriate weight (as it is explicit in Eq.(\ref{5.2})).     

Similar relations involve the two other Keldysh amplitudes, as may be found from Eqs.(\ref{4.6})(\ref{4.5})(\ref{5.9}),
\begin{eqnarray}  
\lefteqn{G_F(p_{1R},p_{2A},p_{3R},p_{4A}) } \nonumber \\
&+&N_F(P_1) \ G_R(p_{1A},p_{2A},p_{3R},p_{4A}) +  N_B(P_3) \ G_R(p_{1R},p_{2A},p_{3A},p_{4A}) \nonumber \\
&=& [ \ N_F(P_1)+N_B(P_3) \ ] \ G_{R/A}^+(p_{1A},p_{2R},p_{3A},p_{4R})
\label{5.11}
\end{eqnarray}
 \begin{eqnarray}
\lefteqn{G_F(p_{1R},p_{2R},p_{3A},p_{4A}) } \nonumber \\
&+&N_F(P_1) \ G_R(p_{1A},p_{2R},p_{3A},p_{4A}) +  N_F(P_2) \ G_R(p_{1R},p_{2A},p_{3A},p_{4A}) \nonumber \\
&=& [ \ N_F(P_1)+N_F(P_2) \ ] \ G_{R/A}^+(p_{1A},p_{2A},p_{3R},p_{4R})
\label{5.12}
\end{eqnarray}
In compact (however misleading) notations, with the use of Eq.(\ref{5.6}), Eqs.(\ref{5.10})(\ref{5.11})(\ref{5.12}) read
\begin{eqnarray}
G^F_{raar} +  N_F(P_1) \ G_{R4}+ N_B(P_4) \ G_{R1} \ = \ [N_F(P_1)+N_B(P_4)]\ G^{R/A}_{RAAR}   \nonumber \\
G^F_{rara} +  N_F(P_1) \ G_{R3}+ N_B(P_3) \ G_{R1} \ = \ [N_F(P_1)+N_B(P_3)] \ G^{R/A}_{RARA}   \nonumber \\
G^F_{rraa} +  N_F(P_1) \ G_{R2}+ N_F(P_2) \ G_{R1} \ = \ [N_F(P_1)+N_F(P_2)] \ G^{R/A}_{RRAA}  
\label{5.13}
\end{eqnarray}

\noindent{\bf Other Keldysh 4-point functions} \\
In subsequent papers devoted to the thermal equilibrium case, and with the help of the Mathematica program, authors of Ref.\cite{Ca1} have noticed that there exists linear combinations of the Keldysh 4-point amplitudes which have simpler properties. They call them "physical combinations" because: i) they satisfy simple KMS conditions \cite{Ca4},  ii) they result in decoupled integral equations for ladder-resummed quantities \cite{Ca5}.
Quoting Eqs.(13)(16) of Ref.\cite{Ca4}, these combinations are
\begin{eqnarray}  
\bar{M}_A\ = \ M_A+N_3M_{R4}+N_4M_{R3}  \nonumber \\
\bar{M}_B\ = \ M_B+N_1M_{R3}+N_3M_{R1} \nonumber \\
\bar{M}_C\ = \ M_C+N_2M_{R3}+N_3M_{R2}  \nonumber \\
\bar{M}_D\ = \ M_D+N_2M_{R4}+N_4M_{R2}  \nonumber \\
\bar{M}_E\ = \ M_E+N_1M_{R2}+N_2M_{R1}  \nonumber \\
\bar{M}_F\ = \ M_F+N_1M_{R4}+N_4M_{R1}  
\label{5.14}
\end{eqnarray}
and the KMS conditions are
\begin{eqnarray} 
(N_1+N_2) \ \bar{M}_A \ = \  (N_3+N_4) \ \bar{M}_E^* \nonumber \\
(N_1+N_4) \ \bar{M}_C \ = \  (N_3+N_2) \ \bar{M}_F^* \nonumber \\
(N_1+N_3) \ \bar{M}_D \ = \  (N_2+N_4) \ \bar{M}_B^* 
\label{5.15}
\end{eqnarray}   
The comparison with Eq.(\ref{5.13}) leads to the identification $G_{Ri} = M_{Ri}$ , and
\begin{equation}   
G^F_{raar} \ = \ M_F \ \ , \ \ \ G^F_{rara} \ = \  M_B \ \ , \ \ \ G^F_{rraa} \ = \ M_E 
\end{equation}
and
\begin{equation} 
\bar{M}_F \ = \ (N_1+N_4) \ G_{RAAR}^{R/A} \ \ , \ \  \bar{M}_B \ = \ (N_1+N_3) \ G_{RARA}^{R/A} \ \ , \ \ \bar{M}_E\ = \ (N_1+N_2) \ G_{RRAA}^{R/A}
\end{equation}  
The condition (\ref{5.6})  is the KMS condition (\ref{5.15}), as it was shown in 1992 in Ref.\cite{vW}. Therefore, the "physical combinations" are the 4-point functions of the $R/A$ basis. And the decoupling of integral equations is, in the $R/A$ basis, a consequence of causality. To conclude,  in thermal equilibrium, the Mathematica program has led from the Keldysh basis  to the $R/A$ basis.        

There are other members  of the Keldysh basis for the 4-point functions, and Ref.\cite{Ca4} give their mutual relations in the thermal equilibrium case. In addition to the four $M_{Ri}$ and the six $M_A$ \dots $M_F$, there are  $M_{\alpha} \ ,  \ M_{\beta} \ , \  M_{\gamma} \ , \ M_{\delta}$ , and $M_T$. Using a method similar to the one used in Sec.\ref{se3.4} for the 3-point function $G_E$, it can be shown that the functions $M_{\alpha}$ \dots $M_{\gamma}$ contain a double discontinuity. For example, for $M_{\alpha}$ , the momentum $P_1$ is kept fixed to $p_{1A}$,  and the diagram is cut into three pieces, where two pieces have, each, one external leg, while the third piece possesses the leg $P_1$ and another leg. One sums over the possible cases for the cuts in the diagram (i.e. $P_1$ connected to $P_2$, or $P_3$, or $P_4$). $M_T$ contains a term associated with the diagram's cuts into four pieces, each one with an external leg.    

\subsection{Ward identities between 4- and 3-point functions}
\label{se4.4}
In Ref.\cite{Ca1}, seven Ward identities are written, which relate  seven 4-point functions $G^{\mu \nu}$ to the 3-point functions $G^{\nu}$. The relations are obtained from simple algebraic identities relating one-loop integrands of the 4-point and 3-point functions, once the approximation $P_i \ll K$ is made in the spin part of the integrand.

As it was the case for the Ward identities between 3-point and 2-point functions, these seven Ward identities are rewritings of the unique IT Ward identity, relating $G_{IT}^{\mu \nu}$ to two $G_{IT}^{\nu}$. They just involve different analytical continuations: i) retarded functions are related to retarded ones, ii) a discontinuity in one energy variable is related to a discontinuity in the same variable, with the same weight. This case involves more energy variables and is more challenging that the 3-point case. It is worth the discussion, which is done in the Appendix. For comparison, the Ward identities obeyed by the $R/A$  4-point functions are also written down.

Moreover, the full one-loop 2-photon-2-fermion amplitude is the sum of the two diagrams (a) and (b) of Fig.\ref{fi2}. It is shown in the appendix that the addition of the two diagrams is incorrectly performed in Ref.\cite{Ca1}, so that four, out of the seven Ward identities written for the full amplitude, have to be corrected.

\section{Conclusion}
\label{se5}
This paper has focused  on one-particle-irreducible amplitudes (1PI) in momentum space.
For an $n$-point amplitude, the number of independent amplitudes is $2^{n-1}-1$ in thermal equilibrium, and $2^n-1$ out of equilibrium. These  numbers are respectively 3 and 7 for $n=3$, or 7 and 15 for $n=4$. In thermal equilibrium, the easiest basis is the Retarded-Advanced basis. For $n=3$, there are the 3 retarded functions (the advanced ones are their complex conjugate, i.e. they are not independent functions). For $n=4$, there are 4 retarded ones, and 3 bi-retarded functions. Out of equilibrium, in the Keldysh basis, the retarded amplitudes belong to the basis. The interpretation of the  other members of the basis has remained obscure.

This work has established the general relation that exists in thermal equilibrium between  3-point and 4-point  1PI functions  of the Keldysh basis and analytical continuations of the corresponding Imaginary-Time  (IT) amplitudes. The intermediary has been the functions of the Retarded-Advanced basis, where a well defined  prescription exists for the weight to be attributed to the many possible analytical continuations  of the IT amplitude.

The relation has emphasized that a set of functions that are particular to the Keldysh basis (set II), has to be interpreted in terms of discontinuity in energy in specific channels of the amplitude. As usual, a discontinuity in a given channel is associated with a process where particles are on-shell in a possible intermediate state of that channel. In other terms, the diagram is cut into two pieces. A characteristic property of those Keldysh amplitudes is that the intermediate states' weight differs from the weight as it appears in the retarded amplitudes. Indeed, for a bosonic state, it is "emission$+$absorption" instead of "emission$-$absorption" (and the reverse for a fermionic state). The nice property of these features is that they are expressed in a form that should carry over to the perturbative out-of-local-equilibrium case. It has been shown for the one-hard-loop case, and it is expected to be general. Indeed, those 1PI amplitudes' properties are likely to be a consequence of the following result. The (connected+disconnected) Green functions in the Keldysh basis have been generally shown by Chou et al \cite{Chou} to be an expectation value of time-ordered nested commutators and anticommutators.

In more precise terms, the external momenta $P_j \ (p_j^0+i\epsilon_j \ , \ {\bf p}_j)$ have to be either of type $R\  (\epsilon_j>0)$ or of type $A\  (\epsilon_j<0)$. The retarded amplitudes (set I) have one leg of type $R$. The amplitudes that belong to set II have two legs of type $R$, $P_{iR}, P_{jR}$, and they are a sum of discontinuities in one energy variable (i.e. in one channel). The energies that appear in the sum are of two types:  i) either they correspond to the energy of an external line of type $R$, i.e. $P_{iR}$ or $P_{jR}$, \  ii) or, in the 4-point case,  they correspond to an external energy that is left undetermined by the $\epsilon$-prescription on the external lines. If the subenergy $p_i^0+p_j^0$ is called the $s$ channel, then the subenergies are those of the $t$ and $u$ channels.  \\
The QED Ward identities relate functions which belong to the same set. Diagramatic rules allow one to write down the loop's integrands. 

There are, in the Keldysh basis, other functions in addition to the sets I and II.  In the 3-point case, one function belongs to set III,  while four do 
in the 4-point case. A general result is that set III contains a double discontinuity (i.e. diagram cut into three pieces). A pattern emerges  for the Keldysh basis,  in momentum space. For example,
for one-loop amplitudes, the retarded amplitude are linear in the distribution functions $n(K_i)$, the  functions belonging to set II are at most bilinear in $n(K_i)$, those of set III at most trilinear,  where  $K_i$ are the momenta running around the loop. 

\appendix
\section{Ward identities between 4- and 3-point functions : IT versus Keldysh}

Just as for the  3-point amplitude in Sec.\ref{se3.5}, the seven Ward identities written in Ref.\cite{Ca1} have to be interpreted as different analytical continuations of the unique IT Ward identity relating 4-point to 3-point  functions. For the diagram (a) of Fig.\ref{fi2}, the relation is (see Sec.7.1.2 in Ref.\cite{MLB})
\begin{equation}
P_{3\mu} \ G_{IT}^{\mu \nu\  (a)}(P_1,P_2, P_3, P_4) = e\ [ G_{IT}^{\nu}(P_1, P_2, P_3+P_4) -G_{IT}^{\nu}(P_1, P_2+P_3, P_4)]
\label{4.10a}
\end{equation}
For this simple one-loop diagram, it will be helpful to remember the origin of the two terms on the right handside of Eq.(\ref{4.10a}). In the contraction with $P_3^{\mu}$ of diagram (a), either the loop momentum $K_3$ disappears and the adjacent momenta $P_3$ and $P_4$ are glued together, or the loop momentum $K_2$ disappears and the adjacent momenta $P_3$ and $P_2$ are glued together.

\subsection{The retarded functions' case}
\label{seA.1}
For each of the retarded amplitudes $G_{Ri}$ defined in Eq.(\ref{4.2}), the analytical continuation of the identity (\ref{4.10a}) is straightforward. The two external momenta that do not take part in the contraction are spectators, i.e. they keep their prescribed value $R$ or $A$. This is enough to determine the prescription $R$ or $A$  on the variable $P_3+P_j$ because $\epsilon_1+\epsilon_2+\epsilon_3+\epsilon_4 =0$. (See the discussion in Sec.\ref{se4.3.1}). For example
\begin{eqnarray}
\lefteqn{P_{3\mu} \ G_R^{\mu \nu\  (a)}(p_{1A},p_{2R}, p_{3A}, p_{4A} )}  \nonumber \\
&=&e\ [ \ G_R^{\nu}(p_{1A}, p_{2R}, (p_3+p_4)_A) - G_R^{\nu}(p_{1A}, (p_2+p_3)_R, p_{4A}) \ ] 
\label{4.10b}
\end{eqnarray}        
\begin{eqnarray}
\lefteqn{P_{3\mu} \ G_R^{\mu \nu\  (a)}(p_{1A},p_{2A}, p_{3R}, p_{4A} ) } \nonumber \\
 &=&e\ [ \ G_R^{\nu}(p_{1A}, p_{2A}, (p_3+p_4)_R) - G_R^{\nu}(p_{1A}, (p_2+p_3)_R, p_{4A})  \ ]
\label{4.10c}
\end{eqnarray}
These equations are Eqs.(67b)(67c) of Ref.\cite{Ca1}. Equations for $G_{R1}^{(a)}$ and  $G_{R4}^{(a)}$ are similar, and they agree with Eqs.(67a)(67d) of Ref.\cite{Ca1}.
\\

\noindent {\bf A comment about the addition of diagrams (a) and (b)} \\
The full 2-photon-2-fermion one-loop amplitude is the sum of the two diagrams (a) and (b) drawn on Fig.\ref{fi2}. Diagram (b) is obtained from diagram (a) by interchanging the external legs $P_3$ and $P_4$, keeping everything else fixed (now $P_3$ is adjacent to $P_4$ and to $P_1$). It is well known that the result of the contraction with $P_{3\mu}$ of diagram (b) is such that the 3-point amplitude $G^{\nu}(P_{1}, P_{2}, P_3+P_4)$ has a sign opposite to the one of diagram (a), so that, in IT formalism, the Ward identity for the sum of the two diagrams is (see Ref.\cite{MLB})
\begin{eqnarray}
\lefteqn{P_{3\mu} \ G_{IT}^{\mu \nu\  (a)+(b)} } \nonumber \\
&=&e\ [ \ G_{IT}^{\nu}((P_{1}+P_3), P_{2}, P_4 ) - G_{IT}^{\nu}(P_{1}, (P_2+P_3), P_{4}) \ ] 
\label{4.11}
\end{eqnarray}
where the term in $(P_1+P_3)$ comes from diagram (b) and the term in $(P_2+P_3)$ from diagram (a).  Following the, above stated, rules for the analytical continuation, it is immediate to write  down the corresponding identity for each of the retarded amplitudes $G_{R1}, G_{R2}, G_{R3}, G_{R4}$. For $G_{R1}$, $G_{R2}$, they agree with Eqs.(69a)(69b) of Ref.\cite{Ca1}, with the correspondence (\ref{2.12})  between compact and explicit notations for the 3-point functions. However, for $G_{R3}$ the analytical continuation give     
\begin{eqnarray}     
\lefteqn{P_{3\mu}G_{R3}^{\mu\nu}  
=P_{3\mu} \ G_{R}^{\mu \nu\  (a)+(b)}(p_{1A},p_{2A},p_{3R}, p_{4A})}  \nonumber \\
&=&e\ [ \ G_{R}^{\nu}((p_{1}+p_3)_R, p_{2A}, p_{4A} ) - G_{R}^{\nu}(p_{1A}, (p_2+p_3)_R,p_{4A}) \ ] 
\label{4.21}
\end{eqnarray}   
which disagrees with the sum of Eqs. (67c) and (68c) of Ref.\cite{Ca1} (where all the subscripts of Eqs.(69c)(69d) are misprinted, since (69c)$\not=$(67c)+(68c) and (69d)$\not=$(67d)+(68d)). If one applies the analytical continuation's rules to their Eqs.(68c)(68d), one can infer from the righthand side that the authors' notations for the lefthand side are
\begin{eqnarray}
G_{R3}^{\mu\nu\ (b)} = G^{\mu\nu\ (b)} (p_{1A}, p_{2A},p_{3A}, p_{4R}) \nonumber \\
G_{R4}^{\mu\nu\ (b)} = G^{\mu\nu\ (b)} (p_{1A}, p_{2A},p_{3R}, p_{4A} )
\label{4.13}
\end{eqnarray}
so that the addition of Eqs.(67c) and (68c) corresponds to
\begin{eqnarray}
G_{R3}^{\mu\nu}=  G^{\mu\nu\ (a)} (p_{1A}, p_{2A},p_{3R}, p_{4A}) + G^{\mu\nu\ (b)} (p_{1A}, p_{2A},p_{3A}, p_{4R})
\label{4.14}
\end{eqnarray}           
and a similar one for $G_{R4}$. Eq.(\ref{4.14}) conflicts with the definition of a retarded amplitude $G_{R3}$. In position space, $t_3$  is always  the latest time, i.e. in momentum space, $p_3$ should be $p_{3R}$ in both terms. (The  retarded amplitude $G_{R3}$ enters the "induced current", i.e. the response of the system to an external (photon) perturbation. See Sec.5.5  in Ref.\cite{BI3}). The correct addition is,  with the correspondence (\ref{4.13}) with the authors' notation,  $$G_{R3}^{\mu\nu\ (a)+(b)} = G_{R3}^{\mu\nu\ (a)} + G_{R4}^{\mu\nu\ (b)}$$
i.e. one should add Eqs.(67c)and (68d) for $G_{R3}$, and Eqs.(67d) and (68c) for $G_{R4}$. The sums now agree with the analytical continuation's results. A similar misidentification occurs in the Ward identities involving the amplitudes $G_F$.

\subsection{The Keldysh amplitudes' case}
\label{seA.2}      
\subsubsection{Diagram (a)}
Turning to the interpretation of the Ward identities involving discontinuities, we first concentrate on diagram (a), i.e. on the  IT Ward identity (\ref{4.10a}). From Eq.(\ref{4.7}), $G_{raar}^{\mu\nu\ (a)}=G_F^{\mu\nu\ (a)}(p_1,p_{2A},p_{3A},p_4)$ is the sum of three discontinuities: in $p_1^0,  \  p_4^0$, and $p_1^0+p_2^0$ while $p_{2A}, p_{3A}$ are being  kept fixed. In the contraction $P_{3\mu}G^{\mu\nu\ (a)}$, there is one term where $P_3$ and $P_2$ are glued, and one term where $P_3$ and $P_4$ are glued:
\\ i) when $P_3$ and $P_2$ are glued, the resulting 3-point amplitude has discontinuities in $p_1^0$, in $p_4^0$,  and it cannot have one in $p_1^0+p_2^0$, which does not exist as a variable in this amplitude. The contraction's  result for this case is, with the identification from Eq.(\ref{2.20b}),
\begin{eqnarray}
N(P_1) \ {\mathrm Disc}_{p_1} G(p_1,(p_2+p_3)_A, p_4) + N(P_4) \ {\mathrm Disc}_{p_4} G(p_1,(p_2+p_3)_A, p_4) \nonumber \\
= \ G_F(p_1,(p_2+p_3)_A,p_4)
\label{4.15}
\end{eqnarray}
 ii)  when $P_3$ and $P_4$ are glued, the resulting 3-point function has a discontinuity in $p_1^0$, a discontinuity in $p_3^0+p_4^0=-(p_1^0+p_2^0)$, while the variable $p_4^0$ does not exists. One obtains, with the identification from Eq.(\ref{2.20b}),
\begin{eqnarray}  
N(P_1) \ {\mathrm Disc}_{p_1} G(p_1,p_{2A}, p_3+p_4) + N(P_3+P_4) \ {\mathrm Disc}_{p_3+p_4} G(p_1,p_{2A}, p_3+ p_4) \nonumber \\
= \ G_F(p_1,p_{2A},p_3+p_4)
\label{4.16}
\end{eqnarray}   
The addition of cases i) and ii), with appropriate signs, is 
\begin{eqnarray}     
P_{3\mu}G_{raar}^{\mu\nu\ (a)} = P_{3\mu}G_F^{\mu\nu\ (a)}(p_1,p_{2A},p_{3A},p_4)  \nonumber \\
= e \ [G_F^{\nu}(p_1,p_{2A}, p_3+p_4)-G_F^{\nu}(p_1,(p_2+p_3)_A,p_4) ]
\label{4.17}
\end{eqnarray}   
Equation (\ref{4.17}) is Eq.(67f) of Ref.\cite{Ca1} with the correspondence (\ref{2.14}) between notations. A similar interpretation occurs for $G_{rraa}^{\mu\nu\ (a)}$.    

From Eq.(\ref{4.9}), the case  $G_{rara}^{\mu\nu\ (a)}=G_F^{\mu\nu\ (a)}(p_1,p_{2A},p_{3},p_{4A})$ has four discontinuities: in $p_1^0,p_3^0,p_1^0+p_4^0,p_1^0+p_2^0$. With $P_3,P_4$ glued, discontinuities exist in $p_1^0$, and  in $p_3^0+p_4^0=-(p_1^0+p_2^0)$. With $P_3,P_2$ glued, discontinuities exist in $p_1^0$, and in $p_2^0+p_3^0=-(p_1^0+p_4^0)$. One obtains
\begin{eqnarray}    
P_{3\mu}G_{rara}^{\mu\nu\ (a)} = P_{3\mu}G_F^{\mu\nu\ (a)}(p_1,p_{2A},p_{3},p_{4A}) \nonumber \\
= e \ [G_F^{\nu}(p_1,p_{2A}, p_3+p_4)-G_F^{\nu}(p_1,(p_2+p_3),p_{4A}) ]
\label{4.18}
\end{eqnarray}           
Equation (\ref{4.18}) is Eq.(67e) of Ref.\cite{Ca1}. The conclusion of the discussion is: the Ward identities relating the 4-point and 3-point Keldysh-specific amplitudes relate discontinuities, keeping track of the allowed variables in the 3-point amplitudes that result from the contraction. 
This feature allows the generalization to the non-equilibrium case.

\subsubsection{Diagram (b)}  
It is worth examining some more examples, i.e. looking at diagram (b) where now $P_3$ is adjacent to $P_4$ and $P_1$. If one considers  $G_F^{\mu\nu\ (b)}(p_1,p_{2A},p_{3A},p_{4})$, one sees that the legs of type $A$, whose energy are being kept fixed, are no more adjacent ones, consequently the amplitude is the sum of four discontinuities: in  $p_1^0,p_4^0,p_1^0+p_3^0,p_1^0+p_2^0$, and one obtains
\begin{eqnarray}
P_{3\mu}G_F^{\mu\nu\ (b)}(p_1,p_{2A},p_{3A},p_{4}) \nonumber \\
= e \ [G_F^{\nu}(p_1+p_3,p_{2A}, p_4)-G_F^{\nu}(p_1,p_{2A}, p_3+p_{4}) ]
\label{4.19}
\end{eqnarray}
For  $G_F^{\mu\nu\ (b)}(p_1,p_{2A},p_{3},p_{4A})$, the adjacent legs $P_2,P_4$ are such that $p_2^0,p_4^0$ are being kept fixed, the amplitude has three discontinuities: in $p_1^0, p_3^0,p_4^0+p_3^0$. In the contraction, when $P_3$ and $P_4$ are glued, the resulting 3-point amplitude has discontinuity in $p_1^0$ and in $p_4^0+p_3^0$; when $P_3$ and $P_1$ are glued, none of the variables  $p_1^0, p_3^0,p_4^0+p_3^0$ occur in the resulting 3-point  amplitude, so that one obtains
\begin{equation}
P_{3\mu}G_F^{\mu\nu\ (b)}(p_1,p_{2A},p_{3},p_{4A})
= -e \ G_F(p_1,p_{2A}, p_3+p_{4}) 
\label{4.20}
\end{equation}
Equations (\ref{4.19}),(\ref{4.20})  are respectively Eqs.(68e),(68f) of Ref.\cite{Ca1}.                

As discussed in Sec.\ref{seA.1} for the retarded amplitudes $G_{R3},G_{R4}$, when adding the contribution of diagrams (a) and (b), the external legs, whose energy is being kept fixed, should be the  same in both terms
\begin{eqnarray}
P_{3\mu}G_F^{\mu\nu\ (a)+(b)}(p_1,p_{2A},p_{3A},p_{4}) \nonumber \\
=P_{3\mu}G_F^{\mu\nu\ (a)}(p_1,p_{2A},p_{3A},p_{4}) + P_{3\mu}G_F^{\mu\nu\ (b)}(p_1,p_{2A},p_{3A},p_{4})
\end{eqnarray} 
i.e. Eqs(67f) and (68e) of Ref.\cite{Ca1} should be added. The result is 
\begin{eqnarray}
P_{3\mu}G_F^{\mu\nu\ (a)+(b)}(p_1,p_{2A},p_{3A},p_{4}) \nonumber \\
= e[G_F^{\nu}(p_1+p_3,p_{2A}, p_4)-G_F^{\nu}(p_1,(p_{2}+p_3)_A, p_{4}) ]
\end{eqnarray}
instead of Eq.(69f), and 
\begin{equation}
P_{3\mu}G_F^{\mu\nu\ (a)+(b)}(p_1,p_{2A},p_{3},p_{4A}) 
= -e \ G_F^{\nu}(p_1,p_2+ p_3,p_{4A}) 
\end{equation}
instead of Eq.(69e) (the term in $p_1+p_3$ comes from diagram (b) and the term in $p_2+p_3$ from diagram (a)).      

\subsection{The Ward identities in the $R/A$ basis}  
For comparison, we quote the Ward identities obeyed by the QED 4-point amplitudes in the $R/A$ basis. Each of the $R/A$ amplitudes is obtained from an analytic continuation of the IT amplitude. The prescription on the external momenta is fixed,  and there is a rule for  the subenergies. For diagram (a), the IT Ward identity (\ref{4.10a}) involves $G_{IT}^{\nu}(P_1,P_2,P_3+P_4)-G_{IT}^{\nu}(P_1,P_2+P_3,P_4)$. In an analytical continuation, there may be  a choice for the  subenergies $p_i^0+p_j^0$ and one has to sum over both possiblities, with the weight prescribed  for the analytical continuations of the IT amplitude  (see Eq.(\ref{5.2})). For example  
\begin{eqnarray}
\lefteqn{P_{3\mu}G_{R/A}^{\mu\nu\ (a)}(p_{1A},p_{2R},p_{3R},p_{4A}) =e \ [-G_R^{\nu}(p_{1A},(p_2+p_3)_R,p_{4A})} \nonumber\\
&+&{{\mathcal N}(p_1+p_2,p_4)\over {\mathcal N}(p_1,p_4)} G_R^{\nu}(p_{1A},p_{2R},(p_3+p_4)_R)
+{{\mathcal N}(p_3+p_4,p_1)\over {\mathcal N}(p_1,p_4)} G_R^{\nu}(p_{1A},p_{2R},(p_3+p_4)_A)]\nonumber \\
& &  
\end{eqnarray}
\begin{eqnarray}
\lefteqn{P_{3\mu}G_{R/A}^{\mu\nu\ (a)}(p_{1A},p_{2R},p_{3A},p_{4R}) =  e\ [ } \nonumber \\
&\  &{{\mathcal N}(p_1+p_2,p_3)\over {\mathcal N}(p_1,p_3)} G_R^{\nu}(p_{1A},p_{2R},(p_3+p_4)_R) \nonumber \\
&+&{{\mathcal N}(p_3+p_4,p_1)\over {\mathcal N}(p_1,p_3)} G_R^{\nu}(p_{1A},p_{2R},(p_3+p_4)_A)\nonumber  \\
&-&{{\mathcal N}(p_1,p_2+p_3)\over {\mathcal N}(p_1,p_3)} G_R^{\nu}(p_{1A},(p_2+p_3)_A, p_{4R})\nonumber \\
&-&{{\mathcal N}(p_1+p_4,p_3)\over {\mathcal N}(p_1,p_3)} G_R^{\nu}(p_{1A},(p_2+p_3)_R,p_{4R})] 
\end{eqnarray}
Similar relations are obtained for $G_{R/A}^{\mu\nu\ (a)}(p_{1A},p_{2A}, p_{3R},p_{4R})$ and for the sum of the diagrams (a) and (b).

\end{document}